\documentclass[12pt]{article}
\usepackage{amssymb,graphicx}

\usepackage{epsf}
\usepackage{graphicx,epsfig}
\usepackage{amsfonts}
\usepackage{amssymb}
\usepackage{cite}

\makeatletter
\renewcommand\section{\@startsection {section}{1}{\z@}%
                                 {-3.5ex \@plus -1ex \@minus -.2ex}
                                   {2.3ex \@plus.2ex}%
                                   {\normalfont\large\bfseries}}
\renewcommand\subsection{\@startsection{subsection}{2}{\z@}%
                                   {-3.25ex\@plus -1ex \@minus -.2ex}%
                                     {1.5ex \@plus .2ex}%
                                     {\normalfont\bfseries}}
\renewcommand\subsubsection{\@startsection{subsubsection}{3}{\z@}%
                                   {-3.25ex\@plus -1ex \@minus -.2ex}%
                                     {1.5ex \@plus .2ex}%
                                     {\normalfont\itshape}}
\makeatother

\newcommand{\Letter}{
\setlength{\textwidth}{16.5cm}
   \setlength{\textheight}{22.6cm}
    \hoffset=-0.6in
\voffset=-2.1cm }

\Letter

\newcommand{\cC}{{\cal C}}

\newcommand{\cI}{{\cal I}}

\newcommand{\RR}{\mathbb{R}}

\renewcommand{\dag}{^{\dagger}}

\newcommand{\rar}{\rightarrow}

\newcommand{\ket}[1]{|#1 \rangle}

\newcommand{\gsim}{ \lower .75ex \hbox{$\sim$} \llap{\raise .27ex \hbox{$>$}} }
\newcommand{\lsim}{ \lower .75ex \hbox{$\sim$} \llap{\raise .27ex \hbox{$<$}} }

\def\RR{R-R}
\def\NSNS{NS-NS}
\def\NSNST{NS-NS,T}
\def\RRT{R-R,T}
\def\s{{\bf s}}
\def\USp{$USp(N_{\alpha}/2)$}
\def\UUSp{$USp(N_{\alpha})$}
\def\U{$U(N_{\alpha}/2)$}

\def\SO{$SO(N_{\alpha}/2)$}
\def\SSO{$SO(N_{\alpha})$}

\begin{document}
\thispagestyle{empty}
\begin{flushright}
\parbox[t]{1.3in}{
MAD-TH-06-1\\
hep-th/0602038}
\end{flushright}

\vspace*{0.5in}

\begin{center}
{\large \bf D-brane Spectrum and K-theory Constraints  \\
\vspace{0.5cm}
of $D=4$, ${\cal N}=1$ Orientifolds}

\vspace*{0.5in} {John Maiden${}^{1}$, Gary Shiu${}^{1}$, and Bogdan
Stefa\'nski, jr.${}^{2}$}
\\[.3in]
{\em
     ${}^1$ Department of Physics,
     University of Wisconsin \\
     Madison, WI 53706, USA \\[.1in]
     ${}^2$ Theoretical Physics Group, Blackett Laboratory,
    Imperial College \\ London SW7 2BZ, U.K.
         }\\[0.5in]
\end{center}

\begin{center}
{\bf
Abstract}
\end{center}
\noindent We study the spectrum of stable BPS and non-BPS D-branes
in ${\bf Z}_2 \times {\bf Z}_2$ orientifolds for all choices of
discrete torsion between the orbifold and orientifold generators. We
compute the torsion K-theory charges in these $D=4$, ${\cal N}=1$
orientifold models directly from worldsheet conformal field theory,
and compare with the K-theory constraints obtained indirectly using
D-brane probes. The K-theory torsion charges derived here provide
non-trivial constraints on string model building. We also discuss
regions of stability for non-BPS D-branes in these examples.

\vfill

\newpage

\section{Introduction}
\label{sec:introduction}

\indent Since their discovery more than a decade
ago~\cite{Polchinski:1995mt}, D-branes have been playing a key role
in elucidating non-perturbative aspects of string theory.
Phenomenologically, they have also become an indispensable tool
because D-branes can localize gauge and matter fields and thus
stable could in fact be where the Standard Model lives. Therefore if
the brane world idea is indeed realized in Nature, it is important
to understand given a string compactification what are the allowed
stable (BPS or non-BPS) D-branes. While the charges of BPS branes
are quite easy to work out, it is in general not a simple task to
enumerate the {\it complete} spectrum of stable D-branes especially
the non-BPS ones except for simple backgrounds such as toroidal
orbifolds~\cite{orbifolds,GabStef,Stef} or
orientifolds~\cite{orientifolds,Sen:1998tt,Frau,StQu}
\footnote{Reviews and further references of the constructions of
these D-brane spectra can be found in~\cite{Sen:1999mg,Gab}.}. As a
result, models whose complete D-brane spectrum has been derived so
far are those with extended supersymmetry and not much is known
about non-BPS branes in the phenomenologically interesting case of
$D=4$, ${\cal N}=1$ supersymmetric backgrounds. In this paper, we
shall address this issue by investigating the spectrum of stable
D-branes for some prototypical  ${\cal N}=1$ examples. In
particular, we will focus on the $\bf{T}^6 / \bf{Z}_2 \times
\bf{Z}_2$ orientifolds because in addition to being simple ${\cal
N}=1$ compactifications, models with realistic particle physics
features have also been constructed in this framework
\cite{Blumenhagen:2005mu}. Moreover, a systematic computer search
for the statistics of supersymmetric D-brane models has  recently
been carried out also for this particular closed string background
\cite{Gmeiner:2005vz, Kumar:2005hf}. Thus, detailed studies of this
specific orientifold  though undoubtedly limited may serve as a
mini-platform for a more ambitious string vacuum project
\cite{SVPwebsite}.

The stability of  a D-brane is typically due to the charges it
carries. Although D-branes were originally discovered as objects
carrying Ramond-Ramond (RR) charges under $p$-form supergravity
fields, their charges are more properly classified by K-theory
\cite{Minasian:1997mm,Witten:1998cd} instead of cohomology. An
important difference between K-theory and cohomology charges arrises
when considering discrete {\it torsion}  (e.g., ${\bf Z}_2$) valued
charges. In fact, the existence of such K-theory torsion charges
(sometimes referred to as K-theory charges) is precisely the reason
that certain non-BPS D-branes are stable \cite{Sen:1999mg}. Due to
the K-theoretical nature of D-brane charges, we expect there are in
general some additional discrete constraints on string constructions
which are invisible in supergravity. Analogously to the usual RR
tadpole conditions, the total torsion charges must cancel in a
consistent string compactification. However, unlike the usual
integral valued RR charges, there are no supergravity fields to
which the torsion charged D-branes are coupled. Hence, the discrete
constraints on the cancellation of torsion charges are invisible
from the usual tadpole conditions obtained by factorization of
one-loop open string amplitudes. Nonetheless, these discrete
K-theory constraints can be detected in an indirect way by
introducing suitable D-brane probes \cite{Uranga:2000xp}. From a
probe brane point of view, a manifestation of these discrete
K-theory constraints is the requirement that there should be an even
number of Weyl fermions charged under the symplectic gauge group on
its worldvolume, for otherwise the worldvolume theory suffers from
global anomalies \cite{Witten:1982fp}. Moreover, it was recently
shown in \cite{Garcia-Etxebarria:2005qc} that for  some specific
simple examples, these discrete constraints from a probe brane
analysis are seen to arise from the standard Dirac quantization
conditions of  4-form fluxes when lifted to F-theory. Although the
probe brane approach provides a powerful way to determine the
K-theory constraints, it is not entirely clear however that all
torsion charges can be obtained in this manner.

We are particularly interested in these K-theory constraints because
they have proven to provide important non-trivial consistency
conditions in building realistic D-brane models from string theory
\cite{GM,Gato-Rivera:2005qd}. Although the K-theory constraints are
automatically satisfied for some simple models
\cite{MarchesanoBuznego:2003hp}, it is certainly not the case in
general. The K-theory constraints of the $\bf{T}^6 / \bf{Z}_2 \times
\bf{Z}_2$ orientifold that we will analyze in detail in this paper
were obtained in \cite{GM} using a probe brane approach. Here, we
would like to determine these K-theory constraints from a direct
conformal field theory (CFT) calculation. It is important to
emphasize that these two approaches are to some extent
complementary. The stability of D-branes and more importantly their
regions of stability are more apparent from the CFT approach, while
the relation to anomaly cancellations is more direct from a probe
brane perspective. By analyzing the one-loop amplitudes for the open
strings that stretch between these branes, as well as imposing
consistency conditions for invariance under the orientifold and
orbifold generators, we can determine the spectrum of stable BPS and
non-BPS branes for a given model. The criteria for a non-BPS brane
to be stable is the absence of tachyonic modes in these open string
amplitudes.

Our results find agreement with the K-theory constraints derived
previously from a probe brane argument \cite{GM}. As we shall see,
the K-theory constraints in \cite{GM} do not constitute the most
general set of K-theory charges but nevertheless they are complete
for the setup considered in \cite{GM}. It will also become clear
later from our spectrum of stable non-BPS branes that for a more
general setup (e.g., when one considers "oblique" magnetic fluxes on
the worldvolumes of D-branes as in
\cite{Antoniadis:2004pp,Bianchi:2005yz}, or D-branes that are stuck
at orbifold fixed points, or D-branes that are not space-filling in
our four dimensional spacetime, etc), then there are further
K-theoretical constraints to be satisfied. In addition to the above
subtleties, there are actually $2^4$ different types of ${\bf Z}_2
\times {\bf Z}_2$ orientifolds corresponding to different choices of
discrete torsion between the orientifold and orbifold
generators~\cite{Klein:2000qw,Braun:2002qa}. The orientifold
background considered in \cite{GM} (which is T-dual to \cite{BL} and
\cite{CSU}) is simply one of them. For completeness, we have also
enumerated the spectra of torsion and integrally charged D-branes
for all other choices of discrete torsion. We expect these results
will be useful for future work in building realistic models from
more general ${\bf Z}_2 \times {\bf Z}_2$ orientifolds.

Finally, another motivation for this work is to explore the
implications of stable D-branes in cosmology. In \cite{Shiu:2003ta},
it was suggested that stable D-branes that are wrapped entirely in
the compact directions (and thus appear pointlike to our
4-dimensional universe) might be interesting cold dark matter
candidates. In particular, the lightest D-particles (LDPs) are
stable because they are the lightest state in the spectrum carrying
a specific charge (either an integral or torsion charge). With the
specific models at hand, we can investigate in a concrete setting
how robust is the existence of such cold dark matter candidates.

This paper is organized as follows. For completeness, we review in
Section \ref{sec:b-torsion} the boundary state method that we use to
compute the spectrum of stable D-brane. Readers who are familiar
with this technique can skip directly to the next section. Section
\ref{sec:setup} gives the specifics of the ${\bf Z}_2 \times {\bf
Z}_2$ orientifolds. To be concrete, we will be working in the T-dual
frame of Type IIB orientifolds with O3 and O7 planes. A table of
orientifold invariant states for two different types of orientifold
projections, known as the hyper-multiplet and the tensor-multiplet
models, is presented. The effects of discrete torsion are also
discussed. To demonstrate explicitly that cancellation of K-theory
charges are additional constraints on top of the usual RR tadpole
cancellations, we have listed the tadpole conditions in Section
\ref{sec:tadpole}. For later comparison with results from the
worldsheet approach, a probe brane analysis of the K-theory
constraints is also included for different  choices of discrete
torsion. A detailed analysis of the torsion brane and integrally
charged brane spectrum is presented in Section \ref{sec:torb}, along
with a discussion of discrete torsion in the model. Some details are
relegated to the appendices. Appendix \ref{sec:amplitudes} contains
the Klein Bottle, M\"{o}bius Strip, and Annulus amplitudes for the
models under consideration. Appendix \ref{sec:tachyon} shows the
calculations used to determine the stability of the non-BPS branes.
Appendix \ref{sec:streg} discusses the stability regions for the
torsion charged branes. Appendix \ref{sec:GPBLcomp} contains the
non-BPS brane spectrum for a ${\bf Z}_2$ orientifold as well as that
of a T-dual version of the ${\bf Z}_2 \times {\bf Z}_2$ orientifold
under consideration. Finally, Appendix \ref{sec:clearing} contains a
table of integrally charged D-matter candidates \cite{Shiu:2003ta}
for different choices of discrete torsion.

\section{Boundary State Formalism}
\label{sec:b-torsion}

In this section we give a brief and elementary review of the
boundary state techniques\footnote{See e.g.
\cite{Polchinski:1987tu,Callan:1987px,DiVecchia:1999rh, GabStef,
Sen, Gab} for a more thorough discussion of the boundary state
method for D-branes.} used in our calculations. This method will
allow us to determine orbifold and orientifold invariant D-branes
states (BPS and non-BPS) in the Type IIB model of \cite{GM},
consisting of O3 and O7 planes, that we want to consider. The
results in the T-dual picture with $O5$ and $O9$ planes can be
obtained by a simple T-duality.

The initial setup for a boundary state calculation is to have a
closed string propagate between two D-brane boundary states, which
are evaluated as matrix elements. When evaluated, this expression
under open/closed string duality gives the familiar one loop
partition function for open strings that end on the D-branes. In
addition, one has to require that the boundary state be invariant
under the closed string GSO projection
($\frac{1}{4}(1+(-1)^{F})(1\pm(-1)^{\tilde{F}})$ where $\mp$
corresponds to IIA/IIB, respectively), as well as the orbifold or
orientifold projection. The $\psi$ coordinates (as well as the
$\partial X$ coordinates) of the superstring are used to define the
boundary state using the condition\footnote{We are using the light
cone gauge, with $x_0$ and $x_9$ as the light cone coordinates.}
\begin{eqnarray} \nonumber
\psi^{\mu}_r+i \eta \tilde{\psi^{\mu}_{-r}} |\eta \rangle = 0 ~~~~~~ \mu = 1,\dots,p+1 \\
\psi^{\mu}_r-i \eta \tilde{\psi^{\mu}_{-r}} |\eta \rangle = 0 ~~~~~~
\mu = p+2,\dots,8
\end{eqnarray}
where $r$ is half-integer moded in the untwisted NSNS sector, and
integer moded in the untwisted RR sector. The boundary state $|\eta
\rangle$ has two values, $\eta = \pm 1$, which correspond to the
different spin structures. In the sectors where $r$ is integer (such
as the untwisted and twisted RR sectors and the ${\bf Z}_2$ twisted
NSNS sector) the ground state is degenerate, giving rise to
additional structure in the boundary state. In this case it is
convenient to define
\begin{eqnarray}
\psi_{\pm}^{\mu}= \frac{1}{\sqrt{2}} (\psi_0^{\mu} \pm i
\tilde{\psi}_0^{\mu})\,,
\end{eqnarray}
which satisfy the usual creation/annihinaltion operator
anti-commutation relations,
\begin{eqnarray}
\{ \psi_{\pm}^{\mu}, \psi_{\pm}^{\nu} \}=0, ~~~~\{ \psi_+^{\mu},
\psi_-^{\nu} \}=\delta^{\mu\nu}.
\end{eqnarray}
In terms of the $\psi_{\pm}$ operators the boundary conditions in
the untwisted RR
sector give
\begin{eqnarray}
\label{eqn:bstate} \nonumber
\psi_{\eta}^{\mu} |\eta \rangle_{\RR} = 0 ~~~~~~ \mu = 1,\dots,p+1 \\
\psi_{-\eta}^{\nu} |\eta \rangle_{\RR} = 0 ~~~~~~ \nu = p+2,\dots,8
\end{eqnarray}
Throughout this paper we will be considering ${\bf Z}_2$ orbifold
actions, and the corresponding ${\bf Z}_2$-twisted sectors. We will
take these ${\bf Z}_2$ actions to invert $n$ spatial coordinates;
the branes we consider will stretch along $s$ of the inverted
directions and $r+1$ un-inverted directions, with $r+s=p$. Since the
${\bf Z}_2$ twisted NSNS sector is integer moded, we also define the
zero mode creation and annihilation operators in this sector, in
terms of which the boundary conditions imply
\begin{eqnarray}
\nonumber
\psi_{\eta}^{\mu} |\eta \rangle_{\NSNST} &=& 0 ~~~~~~ \mu = 9-n,\dots,8-n+s \\
\psi_{-\eta}^{\nu} |\eta \rangle_{\NSNST} &=& 0 ~~~~~~ \nu
=9-n+s,\dots,8
\end{eqnarray}
where we have assumed the orbifold twist acts on $n$ coordinates,
among them $s$ of them are in Neumann directions and $n-s$ are in
Dirichlet directions. The coordinates in the twisted RR sector
behave in a similar manner, but only in the untwisted directions,
\begin{eqnarray}
\nonumber
\psi_{\eta}^{\mu} |\eta \rangle_{\RRT} &=& 0 ~~~~~~ \mu = 1,\dots,r+1 \\
\psi_{-\eta}^{\nu} |\eta \rangle_{\RRT} &=& 0 ~~~~~~ \nu
=r+2,\dots,8-n
\end{eqnarray}
where we have $r+1$ Neumann directions and $7 - n - r$ Dirichlet
directions. Each of our operators (GSO, orientifold, orbifold) can
be written in terms of the $\psi_{\pm}$ operators, which then act on
the boundary states and impose a set of restrictions for the
dimensions of the invariant D-branes.

The action of $\Omega {\cal{I}}_n$ on the fermionic zero modes of a
boundary state\footnote{See the appendix in \cite{StQu} for a
similar treatment of the ${\bf Z}_2$ orientifold
\cite{Pradisi:1988xd,GP}.} is given by
\begin{eqnarray}
\Omega {\cal{I}}_n |\eta \rangle = \kappa \prod_{1}^8 \frac{1 - 2
\psi{}_{0}^i \tilde{\psi}{}_{0}^i}{\sqrt{2}} \prod_{i=9-n}^8
(\sqrt{2} \psi {}_{0}^i) \prod_{i=9-n}^8 (\sqrt{2} \tilde{\psi}
{}_{0}^i) |\eta \rangle
\end{eqnarray}
The first term on the right hand side ($\prod_{1}^8 \frac{1 - 2
\psi{}_{0}^i \tilde{\psi}{}_{0}^i}{\sqrt{2}}$) is the orientifold
action acting on the boundary state $| \eta \rangle$, and is written
as a condition on the zero modes (i.e. when applied to eqn.
(\ref{eqn:bstate}), it takes $\psi_0 \longleftrightarrow
\tilde{\psi_0}$). The other two terms come from the ${\cal{I}}_n$
action, and are also conditions on the zero modes. $\kappa=\pm1$,
and is a phase that allows us to keep our choice of states that are
even or odd under the projection.

Starting with the untwisted sector, our orientifold action is
trivial acting on the NSNS untwisted state, and on the RR
untwisted state we have,
\begin{eqnarray}
\label{eqn:or-stable1} \Omega {\cal{I}}_n |\eta \rangle_{\RR,U} =
\kappa_{\RR,U} i^{5-p+2c+n(n+2)} |\eta \rangle
\end{eqnarray}
$p$ refers to the number of dimensions filled by a D-brane in the
theory, and $c$ is the number of coordinates covered by the $\cI_n$
action that are also filled by the D-brane. $\kappa$ is a phase that
is either $\pm 1$, and is determined by the action of the
orientifold on the untwisted RR sector\footnote{In other words, it
chooses the orientifold projection on the untwisted RR sector to be
symplectic or orthogonal.}.

Next we will approach the twisted sector, which contains an
inherent subtlety that must be explained. When acting on a twisted
sector, the ${\cal{I}}_n$ action does not necessarily transform
the same coordinates as the twisted boundary state (in our case
the twisted sectors created by our orbifold group of 3 ${\bf Z}_2$
generators). When this happens the final result
depends both on the $n$ of the ${\cal{I}}_n$ and the $n'$
coordinates
transformed by the twisted boundary state. We will try to make this
distinction clear in our calculations. For the NSNS twisted sector
we have,
\begin{eqnarray}
\label{eqn:or-stable2} \Omega {\cal{I}}_n |\eta \rangle_{\NSNST} =
\kappa_{\NSNST} i^{2c_c+c_t+n(n+2)+\frac{n'}{2}} |\eta \rangle
\end{eqnarray}
$c_c$ refers to the common filled directions between the $n$ and
$n'$ transformed coordinates. $c_t$ is the number of filled
directions in the twisted boundary state.

In the twisted RR sector we have
\begin{eqnarray}
\label{eqn:or-stable3} \Omega {\cal{I}}_n |\eta \rangle_{\RRT} =
\kappa_{\RRT} i^{5+r+2c_{nc}+n(n+2)-\frac{n'}{2}} |\eta \rangle
\end{eqnarray}
In this case $c_{nc}$ comes from filled compact directions on the
$\bf{T}^6$ that are not common between $n$ and $n'$. $r$ is the
number of filled non-compact directions.

We shall now apply these general calculations to a IIB orientifold
that contains $D9$ branes, which we will use in Section
\ref{sec:setup} when we analyze $D9$ branes with magnetic flux.
Using the equations above, we take $p$ = 9 and so have $c = n$ and
$n' = c_t = 4${}\footnote{$n' = 4$ is due to our ${\bf Z}_2$
generators, which are listed at the end of this section.}. The
results become

\begin{eqnarray}
\Omega {\cal{I}}_n |\eta \rangle_{\RR,U} &=& \kappa_{\RR,U} i^{n(n+4)-4} |\eta \rangle \\
\Omega {\cal{I}}_n |\eta \rangle_{\NSNST} &=& \kappa_{\NSNST} i^{6+2c_c+n(n+2)} |\eta \rangle \\
\Omega {\cal{I}}_n |\eta \rangle_{\RRT} &=& \kappa_{\RRT}
i^{6+2c_{nc}+n(n+2)} |\eta \rangle
\end{eqnarray}

The RR untwisted sector restricts $n$ to be even, which is a result
independent of $p$ for $p$ odd. Thus we can have $n = 0,2,4,6$. With
$n$ even, the twisted sector equations restrict $c_c$ and $c_{nc}$
to be either even or odd. If $c_c$ is even (odd), then $c_{nc}$ must
also be even (odd) relative to the orbifold projections because of
the symmetry of the model. Choosing these to be even, for $n = 2$ or
$6$ we arrive at the orientifold projection in \cite{GM}. For $n =
0$ or $4$, the orientifold projection would be the same as in
\cite{BL}. For our model of $D9$ branes with magnetic flux we will
use the orientifold projection in \cite{GM}, with $n=6$.

Our model will also contain a $\bf{Z}_2 \times \bf{Z}_2$ orbifold
with three projections, which we define to cover the coordinates
defined by the actions $g_1$, $g_2$, and $g_3$, where the $g_i$
orbifold is orthogonal to the $i^{th}$ ${\bf T}^2$:
\begin{eqnarray}
g_1: (z_1,z_2,z_3) &\to& (z_1,-z_2,-z_3), \\
g_2: (z_1,z_2,z_3) &\to& (-z_1,z_2,-z_3), \\ g_3: (z_1,z_2,z_3)
&\to& (-z_1,-z_2,z_3)
\end{eqnarray}
The complex coordinate $z_i$ defines the complex
coordinates\footnote{The complex coordinates $(z_1,z_2,z_3)$
correspond to $z_n = x^{2n+1} + i x^{2n+2}$ where $n=1,2,3$.} on
the $i^{th}$ ${\bf{T}}^2$.

\section{The Setup}
\label{sec:setup} \setcounter{equation}{0}

Having shown how we use the boundary state method to determine
invariance under the orientifold projection, we will now apply these
results to a specific model. After choosing the orientifold
projection in the next section, Section \ref{subsec:GI} lists the
requirements for invariance in each of the untwisted and twisted
NSNS and RR sectors. The section concludes with a brief review of
discrete torsion.

\subsection{Model Specifics}

Now to the specifics of the model. Starting in Type IIA with the
orientifold action in \cite{CSU}, we T-dualize along the $x^3$,
$x^5$, and $x^7$ directions. The result in Type IIB is the
orientifold action $\Omega R$, with $R$ an orbifold projection that
inverts all of the coordinates in the ${\bf{T}}^6$. Due to T-duality
the orientifold action picks up a factor of $(-1)^{F_L}$
\cite{Sen:1996na}. Pairing our orientifold action with our orbifold
generators we obtain 1 O3 and 3 O7 planes, which wrap the
coordinates in the internal directions:
\begin{eqnarray}
\Omega R (-1)^{F_L}:  &&(z_1,z_2,z_3) \to (-z_1,-z_2,-z_3), \\
\Omega R g_1 (-1)^{F_L}: &&(z_1,z_2,z_3) \to (-z_1,z_2,z_3), \\
\Omega R g_2 (-1)^{F_L}: &&(z_1,z_2,z_3) \to (z_1,-z_2,z_3), \\
\Omega R g_3 (-1)^{F_L}:  &&(z_1,z_2,z_3) \to (z_1,z_2,-z_3)
\end{eqnarray}
Finally we come to the branes themselves. The branes fill $r$ + 1
coordinates in the uncompactified space. Since each brane will be
wrapped on a ${\bf T}^2$ in the compactified space, we shall use
$s_i$ as the coordinates on the $i^{th}~{\bf T}^2$. Thus each $s_i$
will be 0, 1, or 2 and a $Dp$ brane will have $p = r + \sum_1^3
s_i$.
From now on we will refer to the dimensions that the branes
fill using the notation $(r;s_1,s_2,s_3)$.

\subsection{Orientifold Invariance}
\label{subsec:GI}

Now that we have chosen the orientifold projection of the model, we
can use an adaptation of equations (\ref{eqn:or-stable1}),
(\ref{eqn:or-stable2}), and (\ref{eqn:or-stable3}) to determine
invariance of the D-brane states under the GSO and orientifold
projections. For Type IIB, the GSO projection requires odd values of
$p$ for $Dp$-branes. Using the boundary state method the invariant
orientifold states obey the restrictions

\begin{eqnarray}
\Omega R (-1)^{F_L}|\eta \rangle_{\RR,U} &=&
\kappa_{\RR,U}^{\Omega} i^{5-p} |\eta \rangle_{\RR,U} \\
\Omega R (-1)^{F_L}|\eta \rangle_{\NSNST_{g_i}} &=&
\kappa_{\NSNST}^{\Omega} i^{s_j+s_k+2} |\eta \rangle_{\NSNST_{g_i}} \\
\Omega R (-1)^{F_L} |\eta \rangle_{\RRT_{g_i}} &=&
\kappa_{\RRT}^{\Omega} i^{5+r+s_i-2} |\eta \rangle_{\RRT_{g_i}}
\end{eqnarray}
Each of the $\kappa$ is a choice of discrete torsion between the
orientifold and the generators of the orbifold group
$\{1,g_1,g_2,g_3\}$. We shall discuss discrete torsion more
thoroughly in the next subsection, but before we can continue, we
need to introduce some terminology. Although our results deal with
4D ${\bf Z}_2 \times {\bf Z}_2$ orientifolds, we are going to borrow
the terms hyper-multiplet and tensor-multiplet, which are used when
describing 6D ${\bf Z}_2$ orientifolds. These terms refer to
different choices of discrete torsion between the orientifold and
orbifold projections. In 6D ${\bf Z}_2$ orbifolds there are both
twisted sector hyper- and tensor-multiplets. One choice of discrete
torsion between the orientifold projection and the twisted sector
keeps the hyper-multiplets \cite{Pradisi:1988xd, GP}, while the
other keeps tensor-multiplets \cite{Pradisi:1988xd, DabPark, Blum}.
This difference was clarified in \cite{Pol2} using the D-brane
language, and in terms of group cohomology in \cite{Braun:2002qa}. A
similar choice of discrete torsion arises between the orientifold
projection and the orbifold generators in the 4D case. Because of
this we have retained this terminology and call our respective 4D
models with those particular choices of orientifold projection the
hyper- and tensor-multiplet models.

Starting with the hyper-multiplet model~\cite{Pradisi:1988xd,GP} in
the $\bf{T}^6 / \bf{Z}_2 \times \bf{Z}_2$, the orientifold and
orbifold invariant states are

\begin{eqnarray}
\label{eqn:GPorient}
\ket{B(r,\s)}_{\mbox{\scriptsize\NSNS}} &&\mbox{for all $r$ and $s_i$} \\
\ket{B(r,\s)}_{\mbox{\scriptsize\RR}} &&\left\{\begin{array}{l}
\mbox{for $r=-1,3$ and $s_1=s_2=s_3=0$ or} \nonumber\\
\mbox{for $r=-1,3$ and $s_i=0$, $s_j=s_k=2$ or} \nonumber\\
\mbox{for $r=1$ and $s_1=s_2=s_3=2$ or}\nonumber\\
\mbox{for $r=1$ and $s_i=2$, $s_j=s_k=0$ or}\nonumber\\
\mbox{for $r=2$ and $s_1=s_2=s_3=1$}\end{array}\right.\nonumber\\
\ket{B(r,\s)}_{\mbox{\scriptsize\NSNS ,T$g_i$}} &&\mbox{for all $r$ and $s_i$ and for $s_j=0, s_k=2$}\nonumber\\
\ket{B(r,\s)}_{\mbox{\scriptsize\RR ,T$g_i$}}
&&\left\{\begin{array}{l}
\mbox{for $r=-1$ and $s_i=2$ and all $s_j,s_k$ or} \nonumber\\
\mbox{for $r=0$ and $s_i=1$ and all $s_j,s_k$ or} \nonumber\\ 
\mbox{for $r=1$ and $s_i=0$ and all $s_j,s_k$ or} \nonumber\\
\mbox{for $r=3$ and $s_i=2$ and all $s_j,s_k$} \nonumber\\
\end{array}\right. \nonumber
\end{eqnarray}
These choices are consistent with the closed string spectrum of the
model and can be determined by combining the results in \cite{StQu}
and \cite{Stef}.

The tensor multiplet model has the same orientifold invariant
boundary states in the untwisted sectors as the hyper-multiplet
model, with the twisted sectors invariant boundary states given by

\begin{eqnarray}
\label{eqn:DPBZorient} \ket{B(r,\s)}_{\mbox{\scriptsize\NSNS
,T$g_i$}} &&\left\{\begin{array}{l}
\mbox{for all $r$ and $s_i$ and for $s_j=s_k=0$ or} \\
\mbox{for all $r$ and $s_i$ and for $s_j=s_k=2$} \\
\end{array}\right. \\
\ket{B(r,\s)}_{\mbox{\scriptsize\RR ,T$g_i$}}
&&\left\{\begin{array}{l}
\mbox{for $r=-1$ and $s_i=0$ and all $s_j,s_k$ or} \nonumber\\
\mbox{for $r=1$ and $s_i=2$ and all $s_j,s_k$ or} \nonumber\\
\mbox{for $r=2$ and $s_i=1$ and all $s_j,s_k$ or} \nonumber\\ 
\mbox{for $r=3$ and $s_i=0$ and all $s_j,s_k$} \nonumber\\
\end{array}\right.
\end{eqnarray}

Given the above restrictions on $r$ and $s_i$ we can construct four
different types of integrally charged branes:

\begin{itemize}
\item Fractional Branes: These are charged under untwisted and
twisted RR forms. There are two types of fractional branes, singly
fractional branes coupling to the $g_i$ twisted sector, which are of
the form
\begin{eqnarray}
\ket{D(r,s)} &=& |B(r,s)>_{\mbox{\scriptsize\NSNS}}
+~|B(r,s)>_{\mbox{\scriptsize\RR}} \\ \nonumber
&&+~|B(r,s)>_{\mbox{\scriptsize$\NSNST_i$}}
+~|B(r,s)>_{\mbox{\scriptsize$\RRT_i$}}
\end{eqnarray}

or totally fractional branes, which are of the form
\begin{eqnarray}
\ket{D(r,s)} &=& |B(r,s)>_{\mbox{\scriptsize\NSNS}}
+~|B(r,s)>_{\mbox{\scriptsize\RR}} \\ \nonumber &&+~\sum_{i=1}^3~\{
|B(r,s)>_{\mbox{\scriptsize$\NSNST_i$}}
+~|B(r,s)>_{\mbox{\scriptsize$\RRT_i$}}\}
\end{eqnarray}

The totally fractional branes only exist in the tensor multiplet
model, and are the $(-1;0,0,0)$, $(3;0,0,0)$, and $(1;2,2,2)$
branes. Singly fractional branes exist for $(r,s_i)$ =
$(-1,2),(1,0),(3,2)$ and $(s_j,s_k)$ = $(2,0),(0,2)$ (hyper), or
$(r,s_i)$ = $(-1,0),(1,2),(3,0)$ and $(s_j,s_k)$ = $(0,0),(2,2)$
(tensor).
Note that this includes the tadpole cancelling $D7$
branes.

\item Bulk Branes: These are charged only under the untwisted RR
forms, and are of the form
\begin{eqnarray}
\ket{D(r,s)} &=& |B(r,s)>_{\mbox{\scriptsize\NSNS}}
+~|B(r,s)>_{\mbox{\scriptsize\RR}}\,.
\end{eqnarray}

These exist for ($r=(-1,3), s=0$), and $(r,s)=(1,6)$ (hyper)
or $(r=1; s_i=0, s_j=s_k=1)$ and $(r=3; s_i=2, s_j=s_k=1)$ (tensor).

\item Truncated Branes: These are charged only under the twisted
RR forms. These exist for the invariant states listed in equations
(\ref{eqn:GPorient}) or (\ref{eqn:DPBZorient}), provided that no
fractional branes exist with the same $r$ and $s_i$. We discuss them
in more detail in section~(\ref{subsec:intcharbrane}).

\item Stuck Branes: These branes are not charged under the twisted
RR forms, but are different from bulk branes in that they cannot
move from the fixed points. Before orientifolding such branes
are a pair of fractional
branes with opposite twisted charges; the orientifold projection
removes the moduli which allow the brane to move off the fixed
points.
We will discuss them in Section~\ref{sec:torb}.
\end{itemize}

With the exception of the truncated branes, the above branes are
BPS. The conditions in (\ref{eqn:GPorient}) and
(\ref{eqn:DPBZorient}) guarantee that the D-branes are both orbifold
and orientifold invariant.

\subsection{Discrete Torsion}
\label{subsec:distors}

Discrete torsion in orbifolds \cite{Vafa:1986wx, Vafa:1994rv} has
been studied extensively, especially in the ${\bf Z}_2 \times {\bf
Z}_2$ case. We will be interested in determining the orbifold
invariant boundary states, hence we need to know the effects of
discrete torsion on the D-brane sector. Here we will summarize the
results\footnote{For more details, e.g. \cite{Stef, Gab2, GabStef,
Douglas:1998xa, Douglas:1999hq, Craps:2001xw}.}. Starting with the
projection operator for the orbifold group
\begin{eqnarray}
P = \frac{1}{| \Gamma |} \sum_{g_i ~\epsilon ~\Gamma} g_i
\end{eqnarray}
where $\Gamma$ is the orbifold group that contains elements $g_i$,
and inserting it into the partition function,
\begin{eqnarray}
Z(q, \overline{q}) = \frac{1}{| \Gamma |} \sum_{g_i, g_j ~\epsilon
~\Gamma} \epsilon (g_i,g_j)
Z(q, \overline{q}; g_i,g_j)
\end{eqnarray}
The partition function can pick up a phase $\epsilon
(g_i,g_j)$ between the elements in the orbifold group. For the ${\bf
Z}_2$ projections
we consider, we have two
choices for the phase: the trivial result $\epsilon (g_i,g_j) = 1$
or the non-trivial result $\epsilon (g_i,g_j) = 1$ if $g_i=g_j$, and
$-1$ otherwise. In our case the orbifold group contains $4$
elements: $\{ 1, g_1, g_2, g_3 \}$. Choosing discrete torsion
between the orbifold generators means that components of our
partition function will be modular invariant up to a phase
definition.

Furthermore, there is a relationship between a IIA theory with
(without) discrete torsion and a IIB theory without (with) discrete
torsion. This effect can be seen by noting that we have two choices
for defining the
fermionic zero mode operators which make up the orbifold elements
$g_i$,
\begin{eqnarray}
g_i &=& \prod (\sqrt{2} \psi_0^i) \prod (\sqrt{2} \tilde{\psi}_0^i) \\
\hat{g}_i &=& \prod (2 \psi_0^i \tilde{\psi}_0^i)
\end{eqnarray}
where we sum over the directions acted on by the ${\bf Z}_2$ twists.
These two
definitions are related by discrete torsion
\begin{eqnarray}
\label{eqn:tors} \hat{g}_i | g_j \rangle = \epsilon (g_i,g_j) g_i |
g_j \rangle
\end{eqnarray}
where $\epsilon (g_i,g_j)$ has the non-trivial definition. This is
equivalent to saying that these two definitions are related by
T-duality. The theory that we define to have discrete torsion ($g_i$
or $\hat{g}_i$) is ambiguous, but each theory is unique in that they
have different Hodge numbers.

Initially we will look at a model with discrete torsion $g_1 |g_2
\rangle = + |g_2 \rangle$, which corresponds to a model with the
Hodge numbers ($h_{11}$, $h_{21}$) = (51,3). The model with $g_1
|g_2 \rangle = - |g_2 \rangle$ corresponds to a model with Hodge
numbers ($h_{11}$, $h_{21}$) = (3,51).

\section{RR Tadpole Conditions}
\label{sec:tadpole}
\setcounter{equation}{0}

In this section, we analyze in detail all the tadpole conditions
arising in the Type IIB orientifolds under consideration. The
purpose of doing this is to illustrate that the K-theory torsion
charges, if uncancelled in a string model, do not show up as the
usual tadpole divergences and hence their cancellation impose
additional constraints. The reason is that unlike the usual
homological RR-charges, there are no supergravity fields to which
the K-theory torsion charges are coupled. Therefore, the presence of
these K-theory torsion charges does not affect the asymptotics of
the Klein bottle, M\"{o}bius strip, and Annulus amplitudes in the
closed string channel, which correspond to the exchanges of light
closed string fields. However, just like the usual RR charges, these
K-theory torsion charges need to be cancelled globally in a
consistent model. We will derive such K-theory constraints for the
${\bf Z}_2 \times {\bf Z}_2$ orientifold using a probe brane
approach in this section. The corresponding derivation using a CFT
approach will be presented in Section \ref{sec:torb}.

\subsection{Homological RR Tadpoles}
\label{subsec:calc}

\indent The tree level interaction for BPS branes on the Annulus,
M\"{o}bius Strip, and Klein bottle, not considering the momentum and
winding sums or Chan-Paton factors, is the same as in the T-dual
case \cite{BL}. Since these do not change, we have included the full
calculations (open and closed string channel) in Appendix
\ref{sec:amplitudes}, with eqn.(\ref{eqn:momwin}) below defining the
momentum and winding on the Klein Bottle (primed) and M\"{o}bius
Strip/Annulus (no prime), respectively. The tilded definitions are
the momentum and winding after a Poisson resummation.
\begin{eqnarray}
\label{eqn:momwin}
{M'}_j &= \sum_{n= - \infty}^{\infty}
e^{\frac{-\pi t n^2}{R_j^2}}, ~~ \qquad {W'}_j &= \sum_{m= -
\infty}^{\infty} e^{-\pi t m^2 R_j^2}
\\ \nonumber
{\widetilde{M}'}_j &= \sum_{s= - \infty}^{\infty} e^{\frac{-\pi
R_j^2 s^2}{t}}, ~~ \qquad  {\widetilde{W}'}_j &= \sum_{r= -
\infty}^{\infty} e^{\frac{-\pi r^2}{R_j^2 t}}
\\ \nonumber
M_j &= \sum_{n= - \infty}^{\infty} e^{\frac{-2\pi t n^2}{R_j^2}}, ~~
\qquad W_j &= \sum_{m= - \infty}^{\infty} e^{-2\pi t m^2 R_j^2}
\\ \nonumber
\widetilde{M}_j &= \sum_{s= - \infty}^{\infty} e^{\frac{-\pi R_j^2
s^2}{2t}}, ~~ \qquad \widetilde{W}_j &= \sum_{r= - \infty}^{\infty}
e^{\frac{-\pi r^2}{2 R_j^2 t}}
\end{eqnarray}

Calculating the RR tadpole for the BPS branes is similar to the
T-dual case.
For the purpose of comparison to the K-theory constraints that
we will introduce in the next subsection, here we list out the
untwisted RR tadpole contribution. The full tadpole, including the
twisted contribution and cross terms between different branes
is well known in the literature
\cite{BL}, and has also been calculated taking into account factors
of discrete torsion \cite{Angelantonj:1999ms, Antoniadis:1999ux}.
For completeness, we summarize these results in
Appendix \ref{sec:amplitudes}. The untwisted tadpole is presented
below, where the first line is the Klein Bottle contribution, the
second line comes from the M\"{o}bius Strip, and the final line is
from the induced $D3$ and $D7$ brane charge on the Annulus.

\begin{eqnarray}
\nonumber
&& v_4  \int dl  \left\{ 32~\left( \frac{1}{v_1 v_2 v_3} + \frac
{v_1 v_2}{v_3} +\frac {v_2 v_3}{v_1} +
\frac {v_1 v_3}{v_2}\right) \right.  \\
&& -2 ~\left(\frac{1}{v_1 v_2 v_3} {\rm{Tr}}~\gamma^T_{\Omega 3,3}
\gamma^{-1}_{\Omega 3,3} + \sum_{i=1}^3 \frac{v_j
v_k}{v_i}~{\rm{Tr}} ~\gamma^T_{\Omega 7_i,7_i}\gamma^{-1}_{\Omega
7_i,7_i}\right) \nonumber
\\
&&+ \frac{1}{32}\left(\frac{1}{v_1 v_2 v_3}~{\rm{Tr}} ~\gamma_{1,3}
\gamma_{1,3}^{-1} + \sum_{i=1}^3 \frac{v_j v_k}{v_i}~{\rm{Tr}}
~\gamma_{1,7_i} \gamma_{1,7_i}^{-1} \right) = 0
\end{eqnarray}
where $i \neq j \neq k$, and the indices run from 1 to 3. For
tadpole cancellation we require
\begin{eqnarray}
\gamma_{\Omega 3,3} &=& + \gamma_{\Omega 3,3}^T \\
\gamma_{\Omega 7_1, 7_3} &=& + \gamma_{\Omega 7_1,7_3}^T \\
\gamma_{\Omega 7_2,7_1} &=& + \gamma_{\Omega 7_2,7_1}^T \\
\gamma_{\Omega 7_3,7_2} &=& + \gamma_{\Omega 7_3,7_2}^T
\end{eqnarray}
The final result is then
\begin{eqnarray}
\frac{1}{32} \int dl~ \left\{ \frac{1}{v_1 v_2 v_3} (32 - n_3)^2 +
\sum_{i=1}^3 \frac{v_j v_k}{v_i} (32 - n_{7_i})^2 \right\} = 0
\end{eqnarray}
This result can also be written in terms of magnetic and wrapping
numbers\footnote{See \cite{Uranga} for a more detailed
explanation.}.
The D-branes we are working with have magnetic
flux, which is quantized according to
\begin{eqnarray}
\frac{m_a^i}{2\pi} \int_{\bf{T}^2_i} F_a^i = n_a^i
\end{eqnarray}
$m_a^i$ is the number of times a D-brane wraps the $i^{th}$
$\bf{T}^2$, and $n_a^i$ is the integer units of flux going through
the $i^{th}$ $\bf{T}^2$. We can use these numbers to describe our
D-branes. For example, a D3 brane ($r$ = 3, $s$ = 0) has magnetic
and wrapping numbers
$[(n_a^1,m_a^1)\times(n_a^2,m_a^2)\times(n_a^3,m_a^3)]=[(1,0)\times(1,0)\times(1,0)]$.
A D5 brane that wraps the first $\bf{T}^2$ would have numbers
$[(n_a^1,m_a^1)\times(1,0)\times(1,0)]$. Our D-branes must be
invariant under the orientifold group, which means we also introduce
image branes with magnetic and wrapping numbers $(n_a^i,-m_a^i)$. In
addition, we shall consider a general setup with $K$ stacks of $N_a$
D-branes. The RR tadpole conditions are now related to cancellation
of the O-plane charge by the D-brane and its image, which we will
not count separately,
\begin{eqnarray}
\sum_a N_a [\Pi_a] + [\Pi_{O3+O7}] =0
\end{eqnarray}
where have 64 O3 planes with $-1/2$ D3 brane charge and 4 O$7_i$
branes with $-$8 D$7_i$ charge that need to be cancelled. The RR
tadpole conditions for a general choice of discrete torsion are
\begin{eqnarray}
\label{eq:D3b}
\sum_{\alpha} N_{\alpha} n^1_{\alpha} n^2_{\alpha} n^3_{\alpha} &=& 16 \kappa_{\Omega R} \\
\label{eq:D71b} \sum_{\alpha} N_{\alpha} n^1_{\alpha} m^2_{\alpha}
m^3_{\alpha} &=& -16 \kappa_{\Omega Rg_1} \\
\label{eq:D72b}
\sum_{\alpha} N_{\alpha} m^1_{\alpha} n^2_{\alpha} m^3_{\alpha} &=& -16 \kappa_{\Omega Rg_2}\\
\label{eq:D73b} \sum_{\alpha} N_{\alpha} m^1_{\alpha} m^2_{\alpha}
n^3_{\alpha} &=&  -16 \kappa_{\Omega Rg_3}
\end{eqnarray}
(\ref{eq:D3b}) refers to the D3 brane, and (\ref{eq:D71b}),
(\ref{eq:D72b}), (\ref{eq:D73b}) are the D$7_i$ branes. The $\kappa$
factors refer to a choice of discrete torsion between the
orientifold and the untwisted RR boundary state ($\kappa_{\Omega
R}$), and the orientifold and the orbifold generators
($\kappa_{\Omega Rg_i}$). The effects of discrete torsion on the
D-brane spectrum will be discussed in more detail in Section
\ref{sec:distorsrev}.

\subsection{K-theoretical RR Tadpoles from Probe Branes}
\label{subsec:krrprobe}

Besides the homological RR tadpole conditions there are the K-theory
torsion constraints which can be found using D-brane probes.
For example, one can introduce probe $D3$ ($r=3,
s=0$) and $D7$ ($r=3, s_i=s_j=2, s_k=0$) branes and demand that
the
number of Weyl fermions on each of the probe brane worldvolume gauge
theory to be even
for otherwise there are $SU(2)$ $D=4$ global gauge
anomalies \cite{Witten:1982fp}.
This is the approach adopted in
\cite{Uranga:2000xp,GM,MarchesanoBuznego:2003hp}

Though we shall prove our results from a CFT approach in the next
section, for the probe brane approach the K-theory constraints in
the hyper-multiplet are

\begin{eqnarray}
\label{eqn:kusual} \sum_{\alpha} N_{\alpha} m^1_{\alpha} m^2_{\alpha} m^3_{\alpha} &\in& 4 {\mathbb{Z}} \\
\sum_{\alpha} N_{\alpha} n^1_{\alpha} n^2_{\alpha} m^3_{\alpha} &\in&  4 {\mathbb{Z}} \\
\sum_{\alpha} N_{\alpha} n^1_{\alpha} m^2_{\alpha} n^3_{\alpha} &\in& 4 {\mathbb{Z}} \\
\label{eqn:kusual4} \sum_{\alpha} N_{\alpha} m^1_{\alpha}
n^2_{\alpha} n^3_{\alpha} &\in& 4 {\mathbb{Z}}
\end{eqnarray}
which requires an even number of non-BPS torsion charged $D9$ and
$D5$ ($r=3, s_i=2, s_j=s_k=0$) branes. 
These torsion charged branes are non-BPS D-branes that
couple to the NSNS sector (untwisted and/or twisted) but
not to RR fields. Nevertheless, they can be
stable because of the discrete torsion ${\bf Z}_2$ charge they carried
\cite{orbifolds, GabStef, Stef, orientifolds, Frau, Sen:1998tt,
StQu, Sen:1999mg, Gab}. See Section \ref{subsec:torbrane} for a more
precise and detailed definition.

For different choices of discrete torsion, the probe brane approach
gives the following K-theory constraints for $\kappa=1$,

\begin{eqnarray}
\label{eqn:ktors} \sum_{\alpha} N_{\alpha} m^1_{\alpha} m^2_{\alpha}
m^3_{\alpha} \in 4 {\mathbb{Z}} &&\left\{\begin{array}{c} \mbox{for
$\kappa$ + $\kappa_{\Omega Rg_1}$ + $\kappa_{\Omega Rg_2}$ +
$\kappa_{\Omega Rg_3}$ =} \\
\mbox{2 or 4} \end{array}\right. \\ \label{eqn:ktors2} \sum_{\alpha}
N_{\alpha} m^i_{\alpha} n^j_{\alpha} n^k_{\alpha} \in 4 {\mathbb{Z}}
&&\left\{\begin{array}{c} \mbox{for $\kappa$ + $\kappa_{\Omega R}$ +
$\kappa_{\Omega Rg_j}$ + $\kappa_{\Omega
Rg_k}$ =} \\
\mbox{2 or 4,} \\ \mbox{i $\neq$ j $\neq$ k}
\end{array}\right.
\end{eqnarray}
and for $\kappa=-1$,
\begin{eqnarray}
\label{eqn:ktors3} \sum_{\alpha} N_{\alpha} m^1_{\alpha}
m^2_{\alpha} m^3_{\alpha} \in 8 {\mathbb{Z}} &&\mbox{for $\kappa$ +
$\kappa_{\Omega Rg_1}$ + $\kappa_{\Omega Rg_2}$ + $\kappa_{\Omega
Rg_3}$ = 2}  \\ \label{eqn:ktors4} \sum_{\alpha} N_{\alpha}
m^i_{\alpha} n^j_{\alpha} n^k_{\alpha} \in 8 {\mathbb{Z}}
&&\left\{\begin{array}{c} \mbox{for $\kappa$ + $\kappa_{\Omega R}$ +
$\kappa_{\Omega Rg_j}$ + $\kappa_{\Omega
Rg_k}$ = 2} \\
\mbox{i $\neq$ j $\neq$ k}
\end{array}\right.
\end{eqnarray}
where there are different K-theory constraints for different choices
of discrete torsion between the orbifold generators ($\kappa$),
between the orientifold and the untwisted RR boundary state
($\kappa_{\Omega R}$), and the orientifold and the orbifold
generators ($\kappa_{\Omega Rg_j}$). The gauge groups for the open
string spectrum on the probe branes have been provided for different
values of discrete torsion in Table \ref{tab:gaugegroup}. By
comparing this result with eqns (\ref{eqn:ktors}),
(\ref{eqn:ktors2}), we see that the K-theory constraints exist when
the probe brane gives $USp$ gauge groups. In terms of the magnetic
and wrapping numbers of the branes, we will have a symplectic gauge
group on the brane when the charge class $[\mathbf{Q}_a]$ of a
D-brane is invariant under the $\Omega R$ action. For open strings
that begin and end a brane with a symplectic gauge group, the
K-theory constraints restrict the number of chiral fermions to be
even.

As we shall see in the next section, these are not the entire set of
possible K-theory constraints, but the one relevant to our setup.
The probe brane approach (at least for the types of probe branes
that have been introduced in the literature) gives us the
constraints for branes that fill the non-compact space ($r=3$) and
for backgrounds with non-oblique flux. Our results in the next
section show that there are additional torsion charged branes that
might show up in more general flux backgrounds.

\section{Non-BPS Branes}
\label{sec:torb}
\setcounter{equation}{0}

\subsection{Torsion Branes in the ${\bf T}^6/({\bf Z}_2\times {\bf Z}_2)$}
\label{subsec:torbrane}

The first set of non-BPS branes we are going to analyze in the model
are torsion branes. In this section we consider only the
hyper-multiplet model, and will consider other models (the T-dual of
the ${\bf T}^6/{\bf Z}_2\times {\bf Z}_2$ orientifold as well as a
${\bf T}^4/{\bf Z}_2$ orientifold) in Appendix \ref{sec:GPBLcomp}.
The effects of discrete torsion will be addressed in Section
\ref{sec:distorsrev}. These non-BPS torsion branes do not couple to
the untwisted or twisted RR sector, i.e. there are no $(-1)^F$
factors in the corresponding open string projection operators.
Torsion branes have boundary states of the form
\begin{eqnarray}
\label{eqn:notors} \ket{D(r,s)} &=& |B(r,s)>_{\mbox{\scriptsize\NSNS}} \\
\hbox{or}  \qquad \label{eqn:onetors}  \ket{D(r,s)} &=&
|B(r,s)>_{\mbox{\scriptsize\NSNS}} +~
\epsilon_i~ |B(r,s)>_{\mbox{\scriptsize$\NSNST_i$}} \qquad i=1,2, \mbox{or}~3\\
\hbox{or} \qquad \label{eqn:threetors} \ket{D(r,s)} &=&
|B(r,s)>_{\mbox{\scriptsize\NSNS}} + \sum_{i=1}^3 \epsilon_i~
|B(r,s)>_{\mbox{\scriptsize$\NSNST_i$}}
\end{eqnarray}
where each of the twisted boundary states is defined up to a phase
$\epsilon_i = \pm 1$ and $\epsilon_3 = \epsilon_1 \epsilon_2$.
The open strings living on these branes are, respectively, invariant
under the following projection operators
\begin{eqnarray}
&& \label{eqn:notorsproj} \left(\frac{1+\Omega R}{2}\right) \\
\label{eqn:onetorsproj} && \left(\frac{1+\Omega
R}{2}\right)\left(\frac{1+g_i}{2}\right) \\
\label{eqn:threetorsproj} && \left(\frac{1+\Omega
R}{2}\right)\left(\frac{1+g_1}{2}\right)\left(\frac{1+g_2}{2}\right)
\end{eqnarray}
Imposing the orientifold projection places restrictions on the
allowed $r$ and $s_i$  values for branes in equations
(\ref{eqn:onetors}) and (\ref{eqn:threetors}). Specifically we see
immediately, from equation (\ref{eqn:GPorient}) that no branes of
the type given in eqn (\ref{eqn:threetors}) are orientifold
invariant. We are now ready to compute the spectrum of stable (i.e.
tachyon-free) branes.

The tachyon is extracted from the open string partition function,
and an equation is set up to cancel the tachyon between the
M\"{o}bius strip and Annulus diagrams. Using this technique, there
are eight possible contributions to the tachyon: the untwisted
Annulus diagram, the twisted Annulus diagram (three contributions),
the M\"{o}bius strip diagram for a boundary state/$O3$ crosscap
interaction, and the M\"{o}bius strip diagram for a boundary
state/$O7_i$ crosscap interaction (three contributions). See
Appendix \ref{sec:tachyon}, eqns (\ref{eqn:cylintach}) -
(\ref{eqn:M71tach}) for the relevant calculations.

The condition for the tachyon to cancel is
\begin{eqnarray}
\label{eqn:tachbegone} && \quad 2^4~n^2~ \times \left(1 +
\epsilon_{T_1} +
\epsilon_{T_2} + \epsilon_{T_3}\right) \nonumber \\
&& - 2n~ \sin \left(\frac{\pi}{4}(r-s+1)\right)
-2n~\sin\left(\frac{\pi}{4}( r+s-2s_3-3 )\right)
\nonumber \\
&& - 2n~\sin\left(\frac{\pi}{4}(r+s-2s_1-3) \right) - 2n~
\sin\left(\frac{\pi}{4}(r+s-2s_2-3 )\right) = 0
\end{eqnarray}
where $n$ is the normalization of the boundary state, and must be
solved for when plugging in values of $r$ and $s$. For stable
torsion branes $n$ must be a non-zero positive number. Because we
cannot construct torsion branes that couple to all three twisted
sectors and are orientifold invariant, we have introduced the
parameter $\epsilon_{T_i}$ to determine to which of the twisted
sectors the brane is coupled. $\epsilon_{T_i}$ = 1 if the brane
couples to the $T_i$ twisted NSNS sector, and 0 otherwise.

It is at this point we would like to emphasize that
eqns.~(\ref{eqn:tachbegone}) and (\ref{eqn:tachbegone2})
are conditions for the existence of torsion charged D-branes of certain
dimensions (i.e., values of $r$ and $s_i$) in a background
with a specific choice of
discrete torsion. The existence of such torsion branes
imply the discrete constraints discussed in Subsection \ref{subsec:krrprobe}.
Hence, for
a vaccum configuration (which involves stacks of D-branes)
to be consistent, we need to check that
the discrete conditions (\ref{eqn:ktors}),(\ref{eqn:ktors2}) (for $\kappa=1$)
or (\ref{eqn:ktors3}), (\ref{eqn:ktors4}) (for $\kappa=-1$)
are satisfied.

There are two types of torsion branes to consider: branes that
couple to twisted NSNS sectors and branes that couple only to the
untwisted NSNS sectors. The former is of the form in eqns.
(\ref{eqn:onetors}), which corresponds to the open string projection
operator (\ref{eqn:onetorsproj}). The latter are of the form in
eqns. (\ref{eqn:notors}), and correspond to the projection operator
(\ref{eqn:notorsproj}).

The branes that couple to twisted NSNS sectors and for which the
open string cancels are listed in Table~\ref{tab:g1torstable}. For a
torsion brane to be invariant under all orbifold and orientifold
projection operators, the branes can only couple to one twisted
sector $T_i$, where $i$ refers to the twisted sector generated by
the $g_i$ orbifold action. It is easy to see that the branes listed
in Table~\ref{tab:g1torstable} are orbifold invariant versions of
the torsion branes found in~\cite{StQu}. For example the
$(3;0,2,0)$-brane coupling to the NSNST$g_i$ sector is a $g_2$
invariant combination of $(5,2)$-branes of~\cite{StQu}.

\begin{table}[ht]
\begin{center}
\begin{tabular}{|c|c|c|} \hline
Invariant Twisted States $T_i$ &$(r;s_1, s_2, s_3)$ &$n^2$ \\ \hline
$i=1$ &$(2;0,2,0)^{\dag}$ &$\frac{1}{128}$ \\
        &$(2;0,0,2)^{\dag}$ &$\frac{1}{128}$ \\
        &$(3;0,2,0)$ &$\frac{1}{64}$ \\
        &$(3;0,0,2)$ &$\frac{1}{64}$ \\
        &$(3;1,2,0)^{\dag}$ &$\frac{1}{128}$ \\
        &$(3;1,0,2)^{\dag}$ &$\frac{1}{128}$ \\ \hline
$i=2$ &$(2;0,0,2)^{\dag}$ &$\frac{1}{128}$ \\
        &$(2;2,0,0)^{\dag}$ &$\frac{1}{128}$ \\
        &$(3;0,0,2)$ &$\frac{1}{64}$ \\
        &$(3;2,0,0)$ &$\frac{1}{64}$ \\
        &$(3;0,1,2)^{\dag}$ &$\frac{1}{128}$ \\
        &$(3;2,1,0)^{\dag}$ &$\frac{1}{128}$ \\ \hline
$i=3$ &$(2;0,2,0)^{\dag}$ &$\frac{1}{128}$ \\
        &$(2;2,0,0)^{\dag}$ &$\frac{1}{128}$ \\
        &$(3;0,2,0)$ &$\frac{1}{64}$ \\
        &$(3;2,0,0)$ &$\frac{1}{64}$ \\
        &$(3;0,2,1)^{\dag}$ &$\frac{1}{128}$ \\
        &$(3;2,0,1)^{\dag}$ &$\frac{1}{128}$ \\ \hline
\end{tabular}
\caption{Stable Torsion branes that couple to twisted NSNS sectors.
These are torsion branes of the form in eqn. (\ref{eqn:onetors}).
Branes that are shown to be inconsistent are marked with a dagger.}
\label{tab:g1torstable}
\end{center}
\end{table}

Of the three types of branes listed in Table~\ref{tab:g1torstable},
not all are consistent. Indeed, following the discussion
in~\cite{Witten:1998cd} and~\cite{Frau}, it was argued
in~\cite{StQu} that the $(4,2)$-branes are inconsistent despite
being tachyon-free. T-dualising the results in
Table~\ref{tab:g1torstable}, it is easy to see that the branes with
$r=2$ or with $s_i=1$ for some $i$, are T-dual to $g_1\times g_2$
orbifold invariant $(4,2)$ branes of~\cite{StQu} and hence are
inconsistent. Thus the $D5$ branes are the only allowed torsion
branes with twisted NSNS coupling. For $i=1$, for example, these
would be the $(3;0,2,0)$ and $(3;0,0,2)$ branes. These can be
thought of as $g_2$ invariant images of the $Z_2 \oplus Z_2$ torsion
branes found in \cite{StQu}.

The torsion branes with one twisted coupling in Table
\ref{tab:g1torstable} can also be thought of as the orientifold
invariant bound state of a fractional BPS D-brane and fractional
anti-BPS D-brane from the ${\bf Z}_2\times{\bf Z}_2$ orbifold. This
can be seen from the normalizations of the D-branes. Indeed, the
normalization squared of a fractional BPS $D5$ brane is
$\frac{1}{256}$. The $D5$ torsion brane has a normalization squared
of $2^2 \times \frac{1}{256} = \frac{1}{64}$ confirming that the
torsion charged brane can be seen as a superposition of a BPS
anti-BPS pair of branes with oppositely charged untwisted and
twisted RR sectors.

The second type of torsion brane, branes that only couple to the
untwisted NSNS sector, have boundary states of the form in
equation~(\ref{eqn:notors}). Our results are presented in Table
\ref{tab:numbertorstable}. Of these branes, in IIB the $D4$ and the
$(r=3;s_i=0,s_j=1,s_k=2)$ branes are T-dual to $D6$ branes, and the
$(2;2,2,2)$ brane is T-dual to a $(2;0,0,0)$ brane. Therefore these
branes are T-dual to branes that can be shown to be inconsistent in
Type I.

\begin{table}[ht]
\begin{center}
\begin{tabular}{|c|c|} \hline
$(r;s_1, s_2, s_3)$ &$n^2$ \\ \hline
$r=2, s_i=2, s_j=s_k=0^{\dag}$ &$\frac{1}{32}$ \\
$r=2, s_1=s_2=s_3=2^{\dag}$ &$\frac{1}{32}$ \\
$r=3, s_i=1, s_j=s_k=0$ &$\frac{1}{32}$ \\
$r=3, s_i=s_j=1, s_k=0$ &$\frac{1}{16}$ \\
$r=3, s_i=2, s_j=s_k=0$ &$\frac{1}{16}$ \\
$r=3, s_1=s_2=s_3=1^{\dag}$ &$\frac{1}{8}$ \\
$r=3, s_i=0, s_j=1, s_k=2^{\dag}$ &$\frac{1}{32}$ \\
$r=3, s_i=2, s_j=s_k=1$ &$\frac{1}{16}$ \\
$r=3, s_i=s_j=2, s_k=1$ &$\frac{1}{32}$ \\
$r=3, s_1=s_2=s_3=2$ &$\frac{1}{16}$ \\ \hline
\end{tabular}
\caption{Stable Torsion branes that only couple to the untwisted
NSNS sectors. These are torsion branes of the form in eqn.
(\ref{eqn:notors}). Branes that are shown to be inconsistent are
marked with a dagger.} \label{tab:numbertorstable}
\end{center}
\end{table}

Note that the $(3;0,2,0)$ branes  of the type given in
equation~(\ref{eqn:notors}) and listed in
Table~\ref{tab:numbertorstable} can be thought of as a pair of
torsion branes of the type given in equation~(\ref{eqn:onetors})
with opposite twisted torsion charges. This is in fact a
$g_2$-invariant version of the process discussed in~\cite{StQu}.

Of the branes in Table \ref{tab:numbertorstable}, the $D9$ and the
$(r=3; s_i=2)$ torsion branes correspond to the branes found by a
probe brane argument in \cite{GM}, using $D9$ branes with
non-oblique magnetic flux on their world-volumes. The two other odd
D-branes, an off-diagonal $D5$ and $D7$ brane, correspond to torsion
charged branes in a configuration that includes a more general
D-brane background.

The additional branes found in Table \ref{tab:numbertorstable} do
not introduce extra K-theory constraints other than the ones
obtained from a probe brane argument reviewed in Section
\ref{subsec:krrprobe}. This can be seen as follows. One can see that
the discrete charges carried by the branes in
Table \ref{tab:numbertorstable} are not independent by showing that
they can decay to one another via changing the compactification
moduli. The tachyon cancellation equation (\ref{eqn:tachbegone}) is
a condition for the absence of ground state tachyons. However,
tachyonic momentum/winding modes can develop as we vary the
compactification radii. The stability region of the branes in Table
\ref{tab:numbertorstable} can be found by generalizing
eqn.~(\ref{eqn:tachbegone}) to include the contributions from
momentum and winding modes. Details are given in Appendix
\ref{sec:streg}. For example, one of the stability conditions for a
$r=3$, $s=1$ brane that fills $x^3$ in the compact space is that
$R_4 \geq \frac{1}{\sqrt{2}}$. If we consider this $D4$ to be a pair
of orientifold invariant truncated branes in the $g_2$ or $g_3$
orbifold, for $R_4 \leq \frac{1}{\sqrt{2}}$ each of these branes can
decay into a pair of orientifold invariant $r=3$, $s_1=2$ $D5$
branes. Therefore the $D4$ and the $D5$ have the same torsion
charge.

As a consistency check one can compute the tree level amplitude
between two torsion charged D-branes. If there a tachyon in the
spectrum, then it signals that there is a common $\bf{Z}_2$ charge
between them which causes an instability in the system. We can use
this method to find an appropriate basis for the branes in terms of
their charges. It can be shown that the branes found in the probe
brane argument are a consistent basis for four $Z_2$ charges, and
that all the branes in Table \ref{tab:numbertorstable} are charged
under at least one of these charges.

To enumerate the spectrum of non-BPS torsion charged D-branes, one
should analyze the stability regions (as discussed in Appendix
\ref{sec:streg}) of the candidate torsion branes in Table
\ref{tab:numbertorstable} and make sure that there are no decay
channels by which they can decay to a pathological brane (e.g.,
those that are T-dual to the $D2$ and $D6$ branes in Type I
\cite{Frau,Witten:1998cd}). Details of such analysis can be found in
Appendix \ref{sec:streg}. However, for the purpose of deriving
discrete K-theoretical constraints in string model building, this
kind of analysis would have to be applied with caution. For example,
in \cite{GM}, the carriers of the discrete K-theory charges are
certain BPS bound state of D-branes (which of course must carry also
the usual homological RR charges in order for them to be BPS). The
D-brane system is BPS only for certain choices of compactification
moduli. The analysis of decay channels in Appendix \ref{sec:streg}
typically involves decompactifying the theory and would take the
D-brane system away from their BPS configuration. It is possible
that non-BPS branes carrying a particular discrete K-theory charges
cannot be constructed even though BPS branes carrying such charges
exist. The stability region of the branes is expanded on in Appendix
\ref{sec:streg}, where a similar decay channel analysis for the
torsion brane spectrum in the hyper- and tensor-multiplet model is
also presented.

\subsection{Integrally Charged Branes}
\label{subsec:intcharbrane}

In this sub-section we make some comments outside the main focus of
this paper by investigating integrally charged D-branes in the ${\bf
Z_2}\times {\bf Z_2}$ orientifolds. In addition to torsion branes
that couple only NSNS sectors, in orbifold and orientifold models
one often finds so-called truncated branes that couple to the
twisted RR sectors. In the present type of models, such integrally
charged branes can have boundary states of one of two types.
Firstly, they could be of the form
\begin{eqnarray}
\label{eqn:intform} \ket{D(r,s)} =
|B(r,s)>_{\mbox{\scriptsize\NSNS}} +
|B(r,s)>_{\mbox{\scriptsize$\RRT_i$}}
\end{eqnarray}
which corresponds to the open string projection operator
\begin{eqnarray}
\left(\frac{1+\Omega
R}{2}\right)\left(\frac{1+g_i(-1)^F}{2}\right)\,.
\end{eqnarray}
Such branes are just the $g_j$ invariant version of the branes found
in \cite{StQu}. Alternately, the truncated brane boundary states can
be of the form
\begin{eqnarray}
\label{eqn:mixedform} \ket{D(r,s)} &=&
|B(r,s)>_{\mbox{\scriptsize\NSNS}} +
|B(r,s)>_{\mbox{\scriptsize$\RRT_i$}} \\ \nonumber &&+~
|B(r,s)>_{\mbox{\scriptsize$\NSNST_j$}} +
|B(r,s)>_{\mbox{\scriptsize$\RRT_k$}}
\end{eqnarray}
which corresponds to the open string projection operator
\begin{eqnarray}
\left(\frac{1+\Omega R}{2}\right)\left(\frac{1+g_i(-1)^F}{2}\right)
\left(\frac{1+g_k(-1)^F}{2}\right)
\end{eqnarray}
Using eqns. (\ref{eqn:GPorient}) and (\ref{eqn:DPBZorient}) we see
that this second type of truncated brane are not possible in the
hyper-multiplet model, but are possible in the tensor-multiplet
model.~\footnote{Branes of the form (\ref{eqn:mixedform}) are
consistent with the orientifold projection in the tensor multiplet
model.}

Calculating the spectrum of integrally charged branes is similar to
finding torsion branes. To determine the integrally charged branes,
the tachyon cancellation condition
\begin{eqnarray}
&& \quad 2^4~n^2~ \times \left(1 + \epsilon_{T_1} +
\epsilon_{T_2} + \epsilon_{T_3}\right) \nonumber \\
&& - 2n~ \sin \left(\frac{\pi}{4}(r-s+1)\right)
-2n~\sin\left(\frac{\pi}{4}( r+s-2s_3-3 )\right) \nonumber \\
\nonumber && - 2n~\sin\left(\frac{\pi}{4}(r+s-2s_1-3) \right) - 2n~
\sin\left(\frac{\pi}{4}(r+s-2s_2-3 )\right) = 0\label{tachcancelint}
\end{eqnarray}
is still useful, but now the twisted sector parameter
$\epsilon_{T_i} = -1$ if the brane couples to the respective twisted
RR sector, and 0 otherwise.

A further condition which restricts the allowed values of $r$ and
$s_i$ for these types of non-BPS D-branes is that there cannot be
any fractional branes with the same values for $r$ and $s_i$. Indeed
if fractional branes exist for a given value of $r$ and $s_i$ we may
always consider a pair of them with opposite bulk RR charge and
suitable Wilson lines, in analogy to~\cite{Sen:1998ex}. Such a
combination also carries the required twisted RR charges. For
completness we list such pairs of fractional branes in
Table~\ref{tab:numberintstable}.

Returning to the tachyon cancelling condition~(\ref{tachcancelint})
it is easy to see that in the case of the hyper-multiplet model the
tachyon only cancels for those values of $r$ and $s_i$ for which
fractional branes exist. Since such pairs of fractional branes are
unstable in certain regimes of moduli space, there will have to be
other D-branes into which these fractional branes decay. Such new
D-branes will have boundary states different from the ones in
equations~(\ref{eqn:intform}) and~(\ref{eqn:mixedform}) and we hope
to investigate them in the future.

\begin{table}[ht]
\begin{center}
\begin{tabular}{|c|c|} \hline
Invariant Twisted States $T_i$ &$(r;s_1, s_2, s_3)$  \\ \hline
$i=1$   &$(-1;2,0,2)$ \\
        &$(-1;2,2,0)$ \\
        &$(1;0,0,2)$ \\
        &$(1;0,2,0)$ \\
        &$(3;2,0,2)$ \\
        &$(3;2,2,0)$ \\ \hline
$i=2$   &$(-1;0,2,2)$ \\
        &$(-1;2,2,0)$ \\
        &$(1;0,0,2)$ \\
        &$(1;2,0,0)$ \\
        &$(3;0,2,2)$ \\
        &$(3;2,2,0)$ \\ \hline
$i=3$   &$(-1;0,2,2)$ \\
        &$(-1;2,0,2)$ \\
        &$(1;0,2,0)$ \\
        &$(1;2,0,0)$ \\
        &$(3;0,2,2)$ \\
        &$(3;2,0,2)$ \\ \hline
\end{tabular}
\caption{Stable brane--anti-brane pairs carrying
twisted RR charges} \label{tab:numberintstable}
\end{center}
\end{table}

In general non-BPS configurations of branes with integral charges
have a possible cosmological application as candidates for cold dark
matter \cite{Shiu:2003ta}. Integrally charged branes with $r=0$ are
interesting because they would appear point-like to a 4D observer,
but could fill some of the compactified dimensions. We have listed
some of the branes for different values of discrete torsion in
Appendix \ref{sec:clearing}.

Besides the non-BPS branes we have been considering, there are also
the BPS stuck branes, which were mentioned briefly at the end of
Section \ref{subsec:GI}. The stuck branes can be divided into those
that are stuck at all the fixed points of the orbifold generators,
or only at one set of fixed points. For the first group we have the
branes $(3;0,0,0)$, $(1;2,2,2)$, and $(-1;0,0,0)$. These branes do
not couple to the twisted sectors, and are located at the fixed
points of the $g_1$ and $g_2$ generators. An example of the second
type of stuck brane is the $(1;0,0,2)$ brane. This brane is stuck
under the $g_3$ orbifold generator but not the $g_1$ or $g_2$. Thus
the model contains three types of $(1;0,0,2)$ branes: 1) Singly
fractional branes with $g_1$ twisted couplings (sitting at the $g_1$
fixed points), 2) Singly fractional branes with $g_2$ twisted
couplings (sitting at the $g_2$ fixed points), and 3) Stuck branes
sitting at the $g_3$ fixed points.

Finally, we would also like to point out that the $(2;1,1,1)$ brane,
which might seem to be a bulk brane, is actually formed from a pair
of branes of the form
\begin{eqnarray}
\ket{D_1(r,s)} &=& |B(r,s)>_{\mbox{\scriptsize\NSNS}}
+~|B(r,s)>_{\mbox{\scriptsize\RR}} \\ \nonumber &&+~
|B(r,s)>_{\mbox{\scriptsize\NSNST}}
+~|B(r,s)>_{\mbox{\scriptsize\RRT}} \\ \nonumber \mbox{and} \\
\nonumber \ket{D_2(r,s)} &=& |B(r,s)>_{\mbox{\scriptsize\NSNS}}
+~|B(r,s)>_{\mbox{\scriptsize\RR}} \\ \nonumber &&-~
|B(r,s)>_{\mbox{\scriptsize\NSNST}}
-~|B(r,s)>_{\mbox{\scriptsize\RRT}}
\end{eqnarray}
For a singly fractional $(2;1,1,1)$ brane that couples to the $g_1$
twisted sector, we could place $\ket{D_1(r,s)}$ at position $x^4$
and $\ket{D_2(r,s)}$ at position $-x^4$, where neither fractional
brane is orientifold or orbifold invariant by itself, but is
invariant as a pair. This brane was discussed
in~\cite{Douglas:1998xa,Douglas:1999hq,Gab2}.

\subsection{Discrete Torsion Revisited}
\label{sec:distorsrev}

In this section we consider discrete torsion in the model and the
extra quantum numbers it introduces:

\begin{itemize}

\item We have to choose the discrete torsion between the orbifold generators $g_1 |g_2 \rangle$. Calling this
choice of phase $\kappa$, the choice $\kappa=-1$ has Hodge numbers
$(h_{11}, h_{21}) = (3,51)$, and is the model with discrete torsion,
and $\epsilon=1$ has $(h_{11}, h_{21}) = (51,3)$, and is the model
without discrete torsion.

\item We have factors of discrete torsion coming from the
orientifold and the orbifold group, $\kappa_{\Omega R g_i}$. There
are four choices to be made: $\Omega R (-1)^{F_L}$ acting on the
untwisted \RR {} boundary state\footnote{i.e., defining the O3
projection to be orthogonal or symplectic.}, and acting on each of
the $|g_i \rangle$ boundary states. See equations
(\ref{eqn:GPorient}) and (\ref{eqn:DPBZorient}) in Section
\ref{subsec:GI} for the invariant states.
\end{itemize}

Though these choices of discrete torsion might seem independent,
they are actually related to each other \cite{Antoniadis:1999ux,
Angelantonj:1999ms}, through the equation \cite{Blumenhagen:2005tn},
\begin{eqnarray}
\label{eqn:distorequate} \kappa_{\Omega R} \kappa_{\Omega Rg_1}
\kappa_{\Omega Rg_2} \kappa_{\Omega Rg_3} = \kappa
\end{eqnarray}

In other words there are $2^4$ choices of discrete torsion that are
allowed in these orientifolds. These choices correspond to the
orbifold discrete torsion, the signs of the RR charge of the
O3-plane, and two of the three O7-planes. These sixteen choices have
also been derived in~\cite{Braun:2002qa} using group cohomology
techniques. In~\cite{Braun:2002qa}, the idea of orbifold discrete
torsion~\cite{Vafa:1986wx}, was generalised to orientifolds. In
particular the allowed orientifolds for a given orientifold were
shown to be classified by a generalised group cohomology with local
coefficients. For the models we have been considering in this paper
the orientifold group is $G=\Omega R(-1)^{F_l}\times g_1\times g_2$,
and the relevant group cohomology was found to
be~\cite{Braun:2002qa}
\begin{equation}
H^2(G,{\tilde U}(1))={\bf Z_2}^{\oplus 4}\,,
\end{equation}
which gives exactly the 16 choices discussed above. The elements of
this
group cohomology, $[H]\in H^2(G,{\tilde U}(1))$
can also be used to define twisted K-theories~\cite{Braun:2002qa}
\begin{equation}
K^{[H]}_G(X)\,.
\end{equation}
which classify the allowed D-brane charges for a brane transverse to
$X$.

We may easily extend our calculations from this section to the
remaining 15 orientifolds in this family. In particular, rewriting
equation (\ref{eqn:tachbegone}) for the torsion branes charged under
the twisted \NSNS {} sector and including factors of discrete
torsion we have
\begin{eqnarray} \label{eqn:tachbegone2}
&& \quad 2^4~n^2 \times (1 + \epsilon_{T_1} + \epsilon_{T_2} +
\epsilon_{T_3})  \\ \nonumber && - 2\kappa_{\Omega R} ~n~
\sin(\frac{\pi}{4}(r-s+1)) - 2\kappa_{\Omega R, g_1}
~n~\sin(\frac{\pi}{4}(r+s-2s_1-3 )) \nonumber \\  &&  - 2
\kappa_{\Omega R, g_2} ~n~ \sin(\frac{\pi}{4}(r+s-2s_2-3 )) -
2\kappa_{\Omega R, g_3} ~n~\sin(\frac{\pi}{4}( r+s-2s_3-3 )) = 0
\nonumber
\end{eqnarray}
Our results in the previous sections dealt with $(\kappa,
\kappa_{\Omega R}, \kappa_{\Omega Rg_1}, \kappa_{\Omega Rg_2},
\kappa_{\Omega Rg_3})$ = $(+,+,+,+,+)$ for the hyper model while for
the tensor model the choice is $(-,+,-,-,-)$. We present an analysis
of the tensor-multiplet torsion brane
spectrum in Appendix \ref{subsec:tortens}.

When calculating the torsion brane spectrum, some of the allowed
branes do not fill the non-compact space (i.e. $r \neq 3$). This
means that the simple probe branes introduced in \cite{GM}, which
fill the non-compact space, will not detect these extra torsion
charges branes, which might lead to extra K-theory constraints. For
the original case we considered (i.e. the hyper-multiplet model),
our torsion brane spectrum only included $r=3$ branes, so our
results matched with the probe brane
argument.

To see the relation between the observed torsion charges from the
probe brane approach and the gauge groups on the $D3$ and $D7_i$
branes, we have included a chart of the gauge groups in Table
\ref{tab:gaugegroup}. Comparing these gauge groups to the K-theory
constraints in eqns (\ref{eqn:ktors}) - (\ref{eqn:ktors4}), we see
that the discrete constraint on $D9$-branes is due to a symplectic
group on the probe $D3$-brane, whereas the discrete constraints on
$D5_i$-branes are the results of symplectic groups on the probe
$D7_i$ branes.

\section{Discussion}
\setcounter{equation}{0}

In this paper, we have investigated D-branes in ${\bf Z}_2 \times
{\bf Z}_2$ orientifolds. We are particularly interested in torsion
charged D-branes because they cannot be detected from the usual
(homological) tadpole conditions. Nevertheless, their charges need
to be cancelled in consistent string vacua and hence their existence
imposes non-trivial constraints on model building. Because of the
generality of the results, we expect the constraints derived here
will be useful for future work in building realistic D-brane models
from more general orientifolds. The search for realistic
intersecting/magnetized D-brane models within the framework of ${\bf
Z}_2 \times {\bf Z}_2$ orientifolds has so far been focused mainly
on a particular choice\footnote{See, however,
\cite{Blumenhagen:2005tn}, for model building from ${\bf Z}_2 \times
{\bf Z}_2$ orientifold with a different choice of discrete torsion.}
of discrete torsion (i.e., the hyper-multiplet model whose non-BPS
brane spectrum was discussed in detail in Section \ref{sec:torb})
and moreover with limited types of branes. For example, the set of
branes considered in \cite{GM} included only bulk branes whose
D-brane charges are induced by turning on "non-oblique"  (in the
sense of \cite{Antoniadis:2004pp,Bianchi:2005yz}) magnetic fluxes on
the worldvolume of D9-branes. As the search continues into different
choices of discrete torsion or when more general branes are
included, we would have to go beyond the K-theory constraints in
eqns (\ref{eqn:kusual}) - (\ref{eqn:kusual4}) (and analogously
eqns.~(\ref{eqn:ktors}) - (\ref{eqn:ktors4}) for other choices of
discrete torsion). The CFT approach adopted here not only reproduces
the probe brane results in \cite{GM}, but finds additional torsion
branes that arise from probe branes that are different from the
simple probes usually considered \cite{GM}. For these more general
cases, one would have to check that the discrete K-theory
constraints are indeed satisfied or else the models are
inconsistent.

So far, the derivations of K-theory constraints have been done in a
case by case basis. It would be useful to have a more general or
perhaps an alternative understanding of how these discrete
constraints arise. The probe brane approach introduced in
\cite{Uranga:2000xp} provides a powerful way to derive some (and in
some cases all) of these discrete constraints. Having checked the
torsion brane spectrum in the ${\bf Z_2} \times {\bf Z_2}$
orientifold for all cases of discrete torsion, we have explicitly
shown that the set of bulk\footnote{By bulk branes, we mean branes
that are not stuck at fixed points and so they cannot carry twisted
NSNS or RR charges.} probe branes with symplectic gauge groups that
one can introduce is in one-to-one correspondence with the set of
bulk torsion branes which are space-filling in the four non-compact
dimensions. Recent work \cite{Garcia-Etxebarria:2005qc} has
suggested yet another interesting way to understand the K-theory
constraints in orientifold constructions by uplifting to F-theory. A
non-vanishing magnetic flux on the world-volume of D-branes in Type
IIB can be encoded by the 4-form flux $G_4$ in F-theory. The
K-theory constraints can then be seen to follow from the standard
Dirac quantization conditions on $G_4$. It would be interesting to
see if a similar analysis can be done for $D=4$, ${\cal N}=1$
orientifold backgrounds such as the ones considered here.

The effects of K-theory constraints have recently been explored in
the statistical studies of string vacua
\cite{Gmeiner:2005vz,Gato-Rivera:2005qd}. The significance of these
discrete constraints in reducing the string landscape is somewhat
model dependent. The authors of \cite{Gmeiner:2005vz} investigated
an ensemble of (homological) tadpole cancelling intersecting D-brane
models in the ${\bf Z_2} \times {\bf Z_2}$ orientifold, and found
that considering K-theory constraints reduced the possible brane
configurations by a factor of five. This is in contrast to
\cite{Gato-Rivera:2005qd} which carried out a similar analysis for
an ensemble of Rational Conformal Field Theory (RCFT) orientifolds
with qualitative features of the Standard Model, and found that the
additional K-theory constraints did not significantly reduce the
number of solutions. It would be interesting to revisit
\cite{Gmeiner:2005vz} for other choices of discrete torsion using
the K-theory constraints derived here.

Finally, our results suggest that the spectrum of stable non-BPS
D-branes for $D=4$, ${\cal N}=1$  orientifolds can be quite rich
(e.g. Table \ref{tab:dmatter} for integrally charged non-BPS
branes). As discussed in \cite{Shiu:2003ta}, stable non-BPS branes
that are point-like in our four non-compact dimensions can be
interesting candidates for cold dark matter. An interesting
direction for future investigation would be to compute their
scattering cross sections and see if they can give rise to sharp
signatures that could distinguish them from other cold dark matter
candidates.

\section*{Acknowledgments}
It is a pleasure to thank F. Marchesano for discussions and
comments. JM and GS would like to thank the Perimeter Institute for
Theoretical Physics for hospitality while part of this work was
done. The work of JM and GS was supported in part by NSF CAREER
Award No. PHY-0348093, DOE grant DE-FG-02-95ER40896, a Research
Innovation Award and a Cottrell Scholar Award from Research
Corporation. The work of BS was supported by a Marie Curie
Fellowship and an EPSRC Advanced Fellowship.

\bigskip \bigskip

\newpage
\noindent {\large \bf Appendices}

\appendix
\section{Some details about the amplitudes}
\label{sec:amplitudes}

\setcounter{equation}{0}

\indent In this appendix, we summarize the results  of
 the full Klein bottle,
M\"{o}bius strip, and Annulus amplitudes \cite{BL,
Antoniadis:1999ux, Angelantonj:1999ms} for the 
${\bf Z}_2 \times {\bf Z}_2$ orientifolds\footnote{See, also
\cite{Forste:2000hx} for this and
other ${\bf Z}_M \times {\bf Z}_N$ orientifolds.}. The expressions are more
complete than that in Section \ref{subsec:calc}
in that we also list the twisted sector contribution to the
partition function. We have also included factors of discrete
torsion, see Section \ref{sec:distorsrev} for more details.
Each of these amplitudes
is calculated first in the open channel and then converted into the
closed channel after a Poisson resummation and a modular
transformation. The winding and momentum terms are written out
explicitly in eqn (\ref{eqn:momwin}). For the modular
transformations, the Klein Bottle has $t=1/4l$; for the M\"{o}bius
Strip, $t=1/8l$; and for the Annulus, $t=1/2l$, where $t$ is the
modulus for the open string one loop amplitude and $l$ is the
modulus for the closed string vacuum amplitude. 
\\
\\ {\bf Klein Bottle}
\begin{eqnarray}
\label{eqn:klein} {\cal K} &=&~\frac{1}{16} \frac{V_4}{(2 \pi)^4}
\int_0^\infty \frac{dt}{2t^3} (\frac{{f_3}^8 (e^{-2 \pi t})}{{f_1}^8
(e^{-2 \pi t})} - \frac{{f_4}^8 (e^{-2 \pi t})}{{f_1}^8 (e^{-2 \pi
t})} - \frac{{f_2}^8 (e^{-2 \pi t})}{{f_1}^8 (e^{-2 \pi t})}) \\
\nonumber && \times (\prod_{i=3}^8 W'_i + \prod_{i=3}^6 M'_i
\prod_{j=7,8} W'_j + \prod_{i=5}^8 M'_i \prod_{j=3,4} W'_j +
\prod_{i=3,4,7,8} M'_i \prod_{j=5,6} W'_j)
\\ \nonumber
&& + 16 \times \frac{1}{8} \frac{V_4}{(2 \pi)^4} \int_0^\infty
\frac{dt}{2t^3} (\frac{{f_3}^4 (e^{-2 \pi t}) {f_2}^4 (e^{-2 \pi
t})}{{f_1}^4 (e^{-2 \pi t}) {f_4}^4 (e^{-2 \pi t})} - \frac{{f_2}^4
(e^{-2 \pi t}) {f_3}^4 (e^{-2 \pi t})}{{f_1}^4 (e^{-2 \pi t})
{f_4}^4 (e^{-2 \pi t})} ) \\ \nonumber && \times \left(
\kappa_{\Omega R} \kappa_{\Omega R g_1} (\kappa \prod_{i=3,4} W'_i +
\prod_{i=3,4} M'_i) + \kappa_{\Omega R} \kappa_{\Omega R g_2}
(\kappa \prod_{i=5,6} W'_i + \prod_{i=5,6} M'_i) \right. \\
\nonumber && \left. + \kappa_{\Omega R} \kappa_{\Omega R g_3}
(\kappa \prod_{i=7,8 }W'_i  + \prod_{i=7,8} M'_i ) \right)
\\ \nonumber
&& =2 \frac{V_4}{(2 \pi)^4} \int_0^\infty dl~ (\frac{{f_3}^8 (e^{-2
\pi l})}{{f_1}^8 (e^{-2 \pi l})} - \frac{{f_2}^8 (e^{-2 \pi
l})}{{f_1}^8(e^{-2\pi l})} - \frac{{f_4}^8 (e^{-2 \pi l})}{{f_1}^8 (e^{-2 \pi l})}) \\
\nonumber && \times (\prod_{i=3}^8 \frac{1}{R_i} \tilde{W}'_i +
\prod_{i=3}^6 R_i \tilde{M}'_i \prod_{j=7,8} \frac{1}{ R_j}
\tilde{W}'_j + \prod_{i=5}^8 R_i \tilde{M}'_i \prod_{j=3,4}
\frac{1}{ R_j} \tilde{W}'_j + \prod_{i=3,4,7,8} R_i \tilde{M}'_i
\prod_{j=5,6} \frac{1}{ R_j} \tilde{W}'_j)
\\ \nonumber
&&+ 16 \frac{V_4}{(2 \pi)^4} \int_0^\infty dl~ (\frac{{f_3}^4 (e^{-2
\pi l}) {f_4}^4 (e^{-2 \pi l})}{{f_1}^4 (e^{-2 \pi l}) {f_2}^4
(e^{-2 \pi l})} - \frac{{f_4}^4 (e^{-2 \pi l}) {f_3}^4 (e^{-2 \pi
l})}{{f_1}^4 (e^{-2 \pi l}) {f_2}^4 (e^{-2 \pi l})} ) \\ \nonumber
&&\times \left( \kappa_{\Omega R} \kappa_{\Omega R g_1} (\kappa
\prod_{j=3,4} \frac{1}{ R_j} \tilde{W}'_j + \prod_{i=3,4} R_i
\tilde{M}'_i) + \kappa_{\Omega R} \kappa_{\Omega R g_2}(\kappa
\prod_{j=5,6} \frac{1}{ R_j}
\tilde{W}'_j + \prod_{i=5,6} R_i \tilde{M}'_i) \right. \\
\nonumber && \left. + \kappa_{\Omega R} \kappa_{\Omega R g_3}
(\kappa \prod_{j=7,8} \frac{1}{ R_j} \tilde{W}'_j + \prod_{i=7,8}
R_i \tilde{M}'_i ) \right)
\end{eqnarray}
\\
\\
{\bf M\"{o}bius Strip}
\\
\\ The following are the results on the M\"{o}bius Strip, written as interaction between a crosscap state and
a boundary state for a $(r;s_1,s_2,s_3)$ brane. Note that when
switching from open to closed channel, the $f_3 (iq)$ and $f_4 (iq)$
functions pick up a factor of $e^{i\pi}$. This is due to the
transformation of the functions with imaginary arguments (see, e.g.,
\cite{Frau}). We mention this because the partition function will
gain a factor of sine, and this term will show up when discussing
torsion branes (see Appendix \ref{sec:tachyon} for more details).
The NS sector to the M\"{o}bius Strip contribution is
\begin{eqnarray}
\label{eqn:mstrip} {\cal M} &=& \frac{1}{16}
\frac{V_{r+1}}{(2\pi)^{r+1}} \kappa_{\Omega R} \int_0^\infty
\frac{dt}{2t} (2t)^{-(r+1)/2}
\\ \nonumber && \times \left(\frac{ f_4^{5+r-s} (ie^{-\pi t}) f_3^{3+s-r} (ie^{-\pi
t}) - f_3^{5+r-s} (ie^{-\pi t}) f_4^{3+s-r}(ie^{-\pi
t})}{f_1^{5+r-s} (ie^{-\pi t})
f_2^{3+s-r}(ie^{-\pi t})  2^{-(3+s-r)/2}} \right) \\
\nonumber && \times \left( {\rm{Tr}}~\gamma^T_{\Omega 3,D_p}
\gamma^{-1}_{\Omega 3,D_p} \prod_{i=6-s} W_i \right)
\\ \nonumber
&&+  \sum_{i=1}^3 \frac{1}{16} \frac{V_{r+1}}{(2\pi)^{r+1}}
\kappa_{\Omega R g_i} \int_0^\infty
\frac{dt}{2t} (2t)^{-(r+1)/2} \frac{1}{2^{-(7-r-s+2s_i)/2}}\\
\nonumber && \times \left(\frac{ f_4^{1+r+s-2s_i} (ie^{-\pi t})
f_3^{7-r-s+2s_i} (ie^{-\pi t}) - f_3^{1+r+s-2s_i} (ie^{-\pi t})
f_4^{7-r-s+2s_i}(ie^{-\pi t})} {f_1^{1+r+s-2s_i} (ie^{-\pi t})
f_2^{7-r-s+2s_i}(ie^{-\pi t})}\right) \\
\nonumber && \times \left( {\rm{Tr}}~\gamma^T_{\Omega 7_i,D_p}
\gamma^{-1}_{\Omega 7_i,D_p} \prod_{j=s-s_i} M_j \prod_{k=2-s_i} W_k
\right)
\\ \nonumber &=& \frac{1}{8} \frac{V_{r+1}}{(2\pi)^{r+1}} \kappa_{\Omega R} \int_0^\infty
dl \left( \frac{ e^{i(\pi/4)(s-r-1)} f_3^{5+r-s} (ie^{-2\pi l})
f_4^{3+s-r} (ie^{-2\pi l})}{f_1^{5+r-s} (ie^{-2\pi l})
f_2^{3+s-r}(ie^{-2\pi l}) 2^{-(3+s-r)/2}} \right. \\ \nonumber  &&
\left. -\frac{ e^{-i(\pi/4)(s-r-1)} f_4^{5+r-s} (ie^{-2\pi l})
f_3^{3+s-r}(ie^{-2\pi l}) }{f_1^{5+r-s}
(ie^{-2\pi l}) f_2^{3+s-r}(ie^{-2\pi l}) 2^{-(3+s-r)/2}} \right) \\
\nonumber && \times \left( {\rm{Tr}}~\gamma^T_{\Omega 3,D_p}
\gamma^{-1}_{\Omega 3,D_p} \prod_{i=6-s} \frac{1}{R_i} \tilde{W}_i
\right) \\ \nonumber  && + \sum_{i=1}^3 \frac{1}{8}
\frac{V_{r+1}}{(2\pi)^{r+1}} \kappa_{\Omega R g_i} \int_0^\infty
dl~\frac{1}{2^{-(7-r-s+2s_i)/2}}\\ \nonumber && \left(\frac{
e^{i(\pi/4)(3-r-s+2s_i)} f_3^{1+r+s-2s_i} (ie^{-2\pi l})
f_4^{7-r-s+2s_i} (ie^{-2\pi l})  } {f_1^{1+r+s-2s_i}
(ie^{-2\pi l}) f_2^{7-r-s+2s_i}(ie^{-2\pi l}) } \right. \\
\nonumber && \left. -\frac{
e^{-i(\pi/4)(3-r-s+2s_i)}f_4^{1+r+s-2s_i} (ie^{-2\pi l})
f_3^{7-r-s+2s_i}(ie^{-2\pi l}) } {f_1^{1+r+s-2s_i} (ie^{-2\pi l})
f_2^{7-r-s+2s_i}(ie^{-2\pi l})}\right)\\ \nonumber && \times \left(
{\rm{Tr}}~\gamma^T_{\Omega 7_i,D_p} \gamma^{-1}_{\Omega 7_i,D_p}
\prod_{j=s-s_i} R_j \tilde{M}_j \prod_{k=2-s_i} \frac{1}{R_k}
\tilde{W}_k \right)
\end{eqnarray}
\\
\\
{\bf Annulus}
\\ 
\\ Below is the Annulus contribution to the partition function for strings stretched between the $D3$ and
$D7$ branes, including contributions from twisted sectors and cross
terms.
\begin{eqnarray}
\label{eqn:annulus} \cC &=& \frac{1}{16} \frac{V_{4}}{(2\pi)^{4}}
\int_0^\infty \frac{dt}{2t} (2t)^{-2} \left(\frac{{f_3}^8 (e^{-\pi
t})}{{f_1}^8 (e^{-\pi t})} - \frac{{f_4}^8 (e^{-\pi t})}{{f_1}^8
(e^{-\pi t})} - \frac{{f_2}^8 (e^{-\pi t})}{{f_1} (e^{-\pi t})}
\right)
\\ \nonumber
&& \times \left( {\rm{Tr}} ~\gamma_{1,D_p} \gamma_{1,D_p}^{-1}
\prod_{s} M_i \prod_{6-s} W_j \right)
\\ \nonumber && + \sum_{i=1}^3 \frac{1}{4} \frac{V_{4}}{(2\pi)^{4}} \int_0^\infty \frac{dt}{2t}
(2t)^{-2} \left(\frac{{f_3}^4 (e^{-\pi t}) {f_4}^4 (e^{-\pi
t})}{{f_1}^4 (e^{-\pi t}) {f_2}^4 (e^{-\pi t})} - \frac{{f_4}^4
(e^{-\pi t}) {f_3}^4 (e^{-\pi t})}{{f_1}^4 (e^{-\pi t}) {f_2}^4
(e^{-\pi t})}\right) \\ \nonumber && \times \left( {\rm{Tr}}
~\gamma_{g_i,D_p} \gamma_{g_i,D_p}^{-1} \prod_{s_i} M_j
\prod_{k=2-s_i} W_k \right)
\\ \nonumber
&& + \frac{1}{8} \frac{V_{4}}{(2\pi)^{4}} \int_0^\infty
\frac{dt}{2t} (2t)^{-2} \left(\frac{{f_3}^4 (e^{-\pi t}) {f_2}^4
(e^{-\pi t})}{{f_1}^4 (e^{-\pi t}) {f_4}^4 (e^{-\pi t})} -
\frac{{f_3}^4 (e^{-\pi t}) {f_2}^4 (e^{-\pi t})}{{f_1}^4 (e^{-\pi
t}) {f_4}^4 (e^{-\pi t})} \right)
\\ \nonumber
&& \times \left( \sum_{i=1}^3 {\rm{Tr}} ~\gamma_{1,D3}
\gamma_{1,D7_i}^{-1} \prod_{s_i} W_i \right)
\\ \nonumber
&& + \frac{1}{8} \frac{V_{4}}{(2\pi)^{4}} \int_0^\infty
\frac{dt}{2t} (2t)^{-2} \left(\frac{{f_3}^4 (e^{-\pi t}) {f_2}^4
(e^{-\pi t})}{{f_1}^4 (e^{-\pi t}) {f_4}^4 (e^{-\pi t})} -
\frac{{f_3}^4 (e^{-\pi t}) {f_2}^4 (e^{-\pi t})}{{f_1}^4 (e^{-\pi
t}) {f_4}^4 (e^{-\pi t})} \right)
\\ \nonumber
&& \times \left( \sum_{i=1, i \neq j \neq k}^3 {\rm{Tr}}
~\gamma_{1,D7_i} \gamma_{1,D7_j}^{-1} \prod_{s_k} M_k \right)
\\ \nonumber
&& - \frac{1}{8} \frac{V_{4}}{(2\pi)^{4}} \int_0^\infty
\frac{dt}{2t} (2t)^{-2} \left(\frac{{f_2}^4 (e^{-\pi t}) {f_4}^4
(e^{-\pi t})}{{f_1}^4 (e^{-\pi t}) {f_3}^4 (e^{-\pi t})} \right)
\times \left( \sum_{i=1}^3 {\rm{Tr}} ~\gamma_{g_i,D3}
\gamma_{g_i,D7_i}^{-1} \prod_{s_i} W_i \right)
\\ \nonumber
&& - \frac{1}{8} \frac{V_{4}}{(2\pi)^{4}} \int_0^\infty
\frac{dt}{2t} (2t)^{-2} \left(\frac{{f_2}^4 (e^{-\pi t}) {f_4}^4
(e^{-\pi t})}{{f_1}^4 (e^{-\pi t}) {f_3}^4 (e^{-\pi t})} \right)
\times \left( \sum_{i=1, i \neq j \neq k}^3 {\rm{Tr}}
~\gamma_{g_k,D7_i} \gamma_{g_k,D7_j}^{-1} \prod_{s_k} M_k \right)
\\ \nonumber
&=& \frac{1}{512} \frac{V_{4}}{(2\pi)^{4}} \int_0^\infty dl~
\left(\frac{{f_3}^8 (e^{-2\pi l})}{{f_1}^8 (e^{-2\pi l})} -
\frac{{f_4}^8 (e^{-2\pi l})}{{f_1}^8 (e^{-2\pi l})} - \frac{{f_2}^8
(e^{-2\pi l})}{{f_1}^8 (e^{-2\pi l})} \right)
\\ \nonumber
&& \times \left( {\rm{Tr}} ~\gamma_{1,D_p} \gamma_{1,D_p}^{-1}
\prod_{s} R_i \tilde{M}_i \prod_{6-s} \frac{1}{R_j} \tilde{W}_j
\right)
\\ \nonumber
&& + \sum_{i=1}^3 \frac{1}{32} \frac{V_{4}}{(2\pi)^{4}}
\int_0^\infty dl~ \left(\frac{{f_3}^4 (e^{-2\pi l}) {f_2}^4
(e^{-2\pi l})}{{f_1}^4 (e^{-2\pi l}) {f_4}^4 (e^{-2\pi l})} -
\frac{{f_2}^4 (e^{-2\pi l}) {f_3}^4 (e^{-2\pi l})}{{f_1}^4 (e^{-2\pi
l}) {f_4}^4 (e^{-2\pi l})}\right) \\ \nonumber && \times \left(
{\rm{Tr}} ~\gamma_{g_i,D_p} \gamma_{g_i,D_p}^{-1} \prod_{j=s_i} R_j
\tilde{M}_j \prod_{k=2-s_i} \frac{1}{R_k} \tilde{W}_k \right)
\\ \nonumber
&& + \frac{1}{64} \frac{V_{4}}{(2\pi)^{4}} \int_0^\infty dl~
\left(\frac{{f_3}^4 (e^{-2\pi l}) {f_4}^4 (e^{-2\pi l})}{{f_1}^4
(e^{-2\pi l}) {f_2}^4 (e^{-2\pi l})} - \frac{{f_3}^4 (e^{-2\pi l})
{f_4}^4 (e^{-2\pi l})}{{f_1}^4 (e^{-2\pi l}) {f_2}^4 (e^{-2\pi l})}
\right)
\\ \nonumber
&& \times \left( \sum_{i=1}^3 {\rm{Tr}} ~\gamma_{1,D3}
\gamma_{1,D7_i}^{-1} \prod_{s_i} \frac{1}{R_i} \tilde{W}_i \right)
\\ \nonumber
&& + \frac{1}{64} \frac{V_{4}}{(2\pi)^{4}} \int_0^\infty dl~
\left(\frac{{f_3}^4 (e^{-2\pi l}) {f_4}^4 (e^{-2\pi l})}{{f_1}^4
(e^{-2\pi l}) {f_2}^4 (e^{-2\pi l})} - \frac{{f_3}^4 (e^{-2\pi l})
{f_4}^4 (e^{-2\pi l})}{{f_1}^4 (e^{-2\pi l}) {f_2}^4 (e^{-2\pi l})}
\right)
\\ \nonumber
&& \times \left( \sum_{i=1, i \neq j \neq k}^3 {\rm{Tr}}
~\gamma_{1,D7_i} \gamma_{1,D7_j}^{-1} \prod_{s_k} R_k \tilde{M}_k
\right)
\\ \nonumber
&& - \frac{1}{64} \frac{V_{4}}{(2\pi)^{4}} \int_0^\infty dl~
\left(\frac{{f_4}^4 (e^{-2\pi l}) {f_2}^4 (e^{-2\pi l})}{{f_1}^4
(e^{-2\pi l}) {f_3}^4 (e^{-2\pi l})} \right) \times \left(
\sum_{i=1}^3 {\rm{Tr}} ~\gamma_{g_i,D3} \gamma_{g_i,D7_i}^{-1}
\prod_{s_i} \frac{1}{R_i} \tilde{W}_i \right)
\\ \nonumber
&& - \frac{1}{64} \frac{V_{4}}{(2\pi)^{4}} \int_0^\infty dl~
\left(\frac{{f_4}^4 (e^{-2\pi l}) {f_2}^4 (e^{-2\pi l})}{{f_1}^4
(e^{-2\pi l}) {f_3}^4 (e^{-2\pi l})} \right) \times \left(
\sum_{i=1, i \neq j \neq k}^3 {\rm{Tr}} ~\gamma_{g_k,D7_i}
\gamma_{g_k,D7_j}^{-1} \prod_{s_k} R_k \tilde{M}_k \right)
\end{eqnarray}

\section{Finding the tachyon}
\label{sec:tachyon}

\setcounter{equation}{0}

As mentioned in Section \ref{subsec:torbrane}, there are eight
possible contribution to the tachyon: the untwisted Annulus diagram,
the twisted Annulus diagram (three contributions), the M\"{o}bius
strip diagram for a string exchange between the $Dp$ and the $O3$
plane, and the M\"{o}bius strip diagram for a string exchange
between the $Dp$ and the $O7$ plane (three contributions). Each of
these contributions is listed below, and was originally calculated
in the closed string channel using the boundary state method. The
open string expression was obtained after a Poisson resummation and
a modular transformation.

Since the orbifold action produces NSNS and RR twisted sectors that
give equal and opposite contributions to the amplitudes, the
parameter $\epsilon_i$ has been introduced. $\epsilon_i = 1$ if the
brane couples to the $T_i$ twisted NSNS sector, and $\epsilon_i =
-1$ if the brane couples to the $T_i$ twisted RR sector. $\epsilon_i
= 0$ if the brane does not couple to the $T_i$ twisted sector.

When these terms are expanded, combining the tachyonic modes
($q^{-1}$) from the Annulus and M\"{o}bius Strip diagrams produces
the constraint equation (\ref{eqn:tachbegone}). After each term we
provide the expansion of the tachyonic and massless modes.

\begin{eqnarray}
\label{eqn:cylintach} {\cal{A}}~&=&~ 2^4 N^2_{(r,s),U}
~\frac{\prod_{6-s} R_j}{\prod_s R_i}~ \int_0^\infty \frac{dt}{2t}
(2t)^{-(r+1)/2} \left([\frac{f_3 (e^{-\pi t})}{f_1 (e^{-\pi
t})}]^8 - [\frac{f_2 (e^{-\pi t})}{f_1 (e^{-\pi
t})}]^8 \right) \\ \nonumber && \times \left( \prod_{s} M_i \prod_{6-s} W_j \right) \\
\nonumber && + \sum_{i=1}^3 2^{2+s-s_i}
N^2_{(r,s),T_i}~\frac{\prod_{2-s_i} R_j}{\prod_{s_i}
R_k}~\epsilon_i~ \int_0^\infty \frac{dt}{2t} (2t)^{-(r+1)/2}
\frac{{f_4}^4 (e^{-\pi t}) {f_3}^4 (e^{-\pi t})}{{f_1}^4 (e^{-\pi
t}) {f_2}^4 (e^{-\pi t})}) \\ \nonumber && \times \left( \prod_{s_i}
M_k \prod_{2-s_i} W_j \right)
\\ \nonumber &=& 2^4 N^2_{(r,s),U} ~\frac{\prod_{6-s}
R_j}{\prod_s R_i}~
\int_0^\infty \frac{dt}{2t} (2t)^{-(r+1)/2} (q^{-1} - 8 q^0 + \dots) \times \prod_{s} M_i \prod_{6-s} W_j \\
\nonumber && + \sum_{i=1}^3 2^{2+s-s_i}
N^2_{(r,s),T_i}~\frac{\prod_{2-s_i} R_j}{\prod_{s_i}
R_k}~\epsilon_i~ \int_0^\infty \frac{dt}{2t} (2t)^{-(r+1)/2}
((4q)^{-1}+\dots) \times \prod_{s_i} M_k \prod_{2-s_i} W_j
\end{eqnarray}
\begin{eqnarray}
\label{eqn:m3tach} {\cal{M}}_3~&=&~ 32N_{(r,s),U} N_{O3} \prod_{6-s}
R_j \int_0^\infty \frac{dt}{2t} (2t)^{-(r+1)/2}
\\ \nonumber && \frac{ e^{-i(\pi/4)(s-r-1)} f_4^{5+r-s} (ie^{-\pi
t}) f_3^{3+s-r} (ie^{-\pi t}) - e^{i(\pi/4)(s-r-1)} f_3^{5+r-s}
(ie^{-\pi t}) f_4^{3+s-r}(ie^{-\pi
t})}{f_1^{5+r-s} (ie^{-\pi t}) f_2^{3+s-r}(ie^{-\pi t})}\\
\nonumber && \times \frac{1}{2^{-(3+s-r)/2}} (\prod_{i=6-s} W_i ) \\
\nonumber &=& 32 N_{(r,s),U} N_{O3} \prod_{6-s} R_j \int_0^\infty
\frac{dt}{2t} (2t)^{-(r+1)/2} (2 \sin \frac{\pi}{4}(r-s+1)) (q^{-1}
+ \dots) \times (\prod_{i=6-s} W_i )
\end{eqnarray}
\begin{eqnarray}
\label{eqn:M71tach} {\cal{M}}_{7_i}~&=&~ 4N_{(r,s),U} N_{O7_i}
\frac{\prod_{2-s_i} R_k}{\prod_{s-s_i} R_j} \int_0^\infty \frac{dt}{2t} (2t)^{-(r+1)/2} \\
\nonumber && (\frac{ e^{-i(\pi/4)(3-r-s+2s_i)} f_4^{1+r+s-2s_i}
(ie^{-\pi t}) f_3^{7-r-s+2s_i} (ie^{-\pi t}) }
{f_1^{1+r+s-2s_i} (ie^{-\pi t}) f_2^{7-r-s+2s_i}(ie^{-\pi t})} \\
\nonumber && - \frac{e^{i(\pi/4)(3-r-s+2s_i)} f_3^{1+r+s-2s_i}
(ie^{-\pi t}) f_4^{7-r-s+2s_i}(ie^{-\pi t})}{f_1^{1+r+s-2s_i}
(ie^{-\pi t}) f_2^{7-r-s+2s_i}(ie^{-\pi t})})
\\ \nonumber &&
\times \frac{1}{2^{-(7-r-s+2s_i)/2}} ( \prod_{i=s-s_i} M_k
\prod_{2-s_i} W_j) \\ \nonumber &=& 4N_{(r,s),U} N_{O7_i}
\frac{\prod_{2-s_i} R_k}{\prod_{s-s_i} R_j} \int_0^\infty
\frac{dt}{2t} (2t)^{-(r+1)/2} (2 \sin \frac{\pi}{4}(r+s-2s_i-3))
(q^{-1} + \dots)
\\ \nonumber  && \times ( \prod_{i=s-s_i} M_k \prod_{2-s_i} W_j)
\end{eqnarray}
where the normalizations for the Annulus contribution are matched to
the open string one loop diagram,
\begin{eqnarray}
N^2_{(r,s),U} &=& \frac{V_{r+1}}{(2\pi)^{r+1}}~n^2~\frac{\prod_s
R_i}{\prod_{6-s} R_j} \\ N^2_{(r,s),T_k} &=&
2^{4-s+s_i}~\frac{V_{r+1}}{(2\pi)^{r+1}}~n^2~ \frac{\prod_{s_k}
R_i}{\prod_{2-s_k} R_j}
\end{eqnarray}
and the normalization for the M\"{o}bius strip contribution comes
from
the normalizations of the crosscap states and are defined
\begin{eqnarray}
N^2_{O3} &=& \frac{1}{1024}~\frac{V_{4}}{(2\pi)^{4}}~
\prod_{s}\frac{1}{R_j} \\
N^2_{O7_i} &=& \frac{1}{16}~\frac{V_{4}}{(2\pi)^{4}}~
\frac{\prod_{4-s+s_i} R_i}{\prod_{s_i} R_j}
\end{eqnarray}
$n$ is a constant that is determined for different values of
($r,s$). For BPS D-branes that couple to all three twisted
sectors, $n^2 = \frac{1}{256}$.

\section{Stability Region of the Torsion Branes}
\label{sec:streg} \setcounter{equation}{0}
\subsection{Higher Winding and Momentum Modes}

In this section we will calculate the stability region of the
torsion branes in Section \ref{subsec:torbrane}. In that section the
branes were considered stable if the ground state tachyonic mode
vanished. To determine the stability region of these branes, we must
analyze the higher winding and momentum modes, and require that
these not become tachyonic.

For a general $(r,s)$ torsion brane with possible twisted couplings,
the potentially tachyonic modes, with higher winding and momentum
included, are

\begin{eqnarray}
&& \quad 2^4~n^2~ q^{-1}~ \times ~\prod_s M_i \prod_{6-s} W_j +
2^4~n^2~\sum_{k=1}^3\epsilon_k ~q^{-1}~\times
~\prod_{s_k} M_i \prod_{2-s_k} W_j \\
\nonumber && - 2n\kappa_{\Omega R}~q^{-1}~ \sin
\left(\frac{\pi}{4}(r-s+1)\right) \times ~\prod_{6-s} W_j \\
\nonumber && -2n\kappa_{\Omega Rg_1}~q^{-1}~\sin\left(\frac{\pi}{4}(
r+s-2s_1-3 )\right) \times~\prod_{s_2+s_3} M_i \prod_{2-s_1} W_j \\
\nonumber && - 2n\kappa_{\Omega
Rg_2}~q^{-1}~\sin\left(\frac{\pi}{4}(r+s-2s_2-3) \right)
\times~\prod_{s_1+s_3} M_i \prod_{2-s_2} W_j \\ \nonumber && -
2n\kappa_{\Omega Rg_3}~q^{-1}~ \sin\left(\frac{\pi}{4}(r+s-2s_3-3
)\right)
\times~\prod_{s_1+s_2} M_i \prod_{2-s_3} W_j
\end{eqnarray}
where $\epsilon_k=1$ if the brane couples to the $g_k$ NSNS twisted
sector, and 0 otherwise. By expanding the winding and momentum
terms, in addition to having the vacuum state cancel (eqn.
(\ref{eqn:tachbegone})), we must restrict the size of the compact
directions in the expression above such that the higher modes are
not tachyonic. For example, the $(3;0,2,0)$ brane that couples to
the $g_1$ twisted NSNS sector, along the $g_1$ twisted directions
$(x^5,x^6,x^7,x^8)$ the brane is stable for
\begin{eqnarray}
\frac{2}{R_{5,6}^2}-1 & \geq & 0 \\
2R_{7,8}^2-1 & \geq & 0\,.
\end{eqnarray}

A $(3;0,2,0)$ torsion brane without any twisted couplings, whose
stability region in any of the compact directions
\begin{eqnarray}
\frac{2}{R_{i}^2}+2R_{j}^2-1 \geq 0\,,\qquad i=5,6\qquad
j=3,4,7,8\,.
\end{eqnarray}

\subsection{Torsion branes in the hyper-multiplet model}
\label{subsec:torhyp}

In Section \ref{sec:torb} we have identified stable, torsion charged
D-branes in the hyper-multiplet model, and have presented in
equation~(\ref{eqn:tachbegone2}) a general condition for the absence
of an open string tachyon on a D-brane for the different allowed
choices of discrete torsion in these models. Enumerating such
tachyon free torsion-charged D-branes is then straightforward. Not
all tachyon free branes however lead to allowed branes in string
theory. In the case of branes that couple to twisted NSNS sectors,
one needs to ensure that the corresponding boundary states are
invariant with respect to the GSO and orientifold projections.
Further, as in the case of Type I $SO$ and $Sp$ theories we need to
ensure that our D-branes do not suffer from the pathologies
associated to the $D6$ and $D2$ branes in those
theories~\cite{Witten:1998cd,Frau}. It is also possible for
apparently consistent D-branes to decay into inconsistent D-branes.
Such decays signal the need to exclude the former
branes~\cite{StQu}. In this sub-section we present the allowed
torsion charged D-branes in the hyper-multiplet model as an example.

For the hyper-multiplet model the choice of discrete torsion is
\begin{equation}
\kappa=1\,,\qquad \kappa_{\Omega R}=1\,,\qquad \kappa_{\Omega R
g_i}=1\,,
\end{equation}
and equation~(\ref{eqn:tachbegone2}) together with the
conditions~(\ref{eqn:GPorient}) imply that tachyon-free torsion
D-branes only exist in the form given in eqns (\ref{eqn:onetors})
and (\ref{eqn:notors}). Torsion branes of the form given in equation
(\ref{eqn:onetors}) with coupling to the NSNST$g_1$ twisted sector
need to have $(r;s_1,s_2,s_3)$ of the form
\begin{equation}
(2;0,2,0)\,,\quad (2;0,0,2)\,,\quad (3;0,2,0)\,,\quad
(3;0,0,2)\,,\quad (3;1,2,0)\,,\quad (3;1,0,2)\,.
\end{equation}
Tachyon-free torsion branes of the form given in equation
(\ref{eqn:notors}) need to have $(r;s_1,s_2,s_3)$ of the form
\begin{eqnarray}
&&(2;2,0,0)\,,\qquad (2;2,2,2)\,,\qquad (3;1,0,0)\,,\qquad
(3;1,1,0)\,,\qquad (3;2,0,0)\,,\qquad \nonumber \\
&&(3;1,1,1)\,,\qquad (3;0,1,2)\,,\qquad (3;2,1,1)\,,\qquad
(3;2,2,1)\,,\qquad (3;2,2,2)\,.
\end{eqnarray}
and all permutations of the $s_i$.

For the torsion branes with one twisted coupling given by equation
(\ref{eqn:onetors}), the $(2;0,2,0)$, $(2;0,0,2)$, $(3;1,2,0)$, and
$(3;1,0,2)$ branes are $g_3$ images of the ${\bf Z_2}\oplus{\bf
Z_2}$ $(4,2)$-branes which were found to be inconsisntent
in~\cite{StQu}. Therefore the only consistent brane with one twisted
coupling is the $(3;2,0,0)$ brane and all permutations of the $s_i$.

When considering possible decay channels, they suggest either one of
two potential decays: a) $Dp$ brane $\rar$ $D(p \pm 1)$ brane or b)
$Dp$ brane $\rar$ $D(p \pm 2)$ brane. General transitions of the
form a) have been beautifully analysed by~\cite{Sen:1998ex}. At a
critical compactification radius the two CFTs that correspond to the
two D-branes are equivalent, allowing a $Dp$ brane to decay into a
$D(p \pm 1)$ brane. Decays of the form b) are in general more
complicated, and it is not clear what are the decay channels allowed
by matching the whole CFTs. For example, while a CFT matching exists
for certain freely acting orbifolds~\cite{MajSen}, in more
complicated settings such as~\cite{Stef,StQu} no such matching
exists and it is in general not known what branes decay into.
Therefore we will only consider decays of the form a) to exclude
inconsistent torsion branes.

Since the torsion branes that couple only to the NSNSU sector (of
the form in equation (\ref{eqn:notors})) are pairs of torsion branes
that couple to the NSNST$g_i$ sector (of the form in equation
(\ref{eqn:onetors})) with opposite twisted torsion charge, then we
must exclude branes that have the same values of $(r;s_1,s_2,s_3)$.
This means that for branes that couple only to the NSNSU sector, the
$(2;2,0,0)$ and $(3;0,1,2)$ are excluded. The $(2;2,2,2)$ and
$(3;1,1,1)$ are T-dual to a $D2$ and $D6$ brane in Type I, and are
also excluded. In addition, for certain values of the
compactification radii there are the allowed decays
\begin{equation}
(3;1,1,0)\rightarrow(3;1,2,0)\,,
\end{equation}
and
\begin{equation}
(3;1,2,1)\rightarrow(3;1,2,0)\,.
\end{equation}
Therefore the consistent stable torsion branes that have no twisted
couplings are
\begin{equation}
(3;1,0,0)\,,\qquad (3;2,0,0)\,,\qquad (3;2,2,1)\,,\qquad
(3;2,2,2)\,.
\end{equation}
and all permutations of the $s_i$.

Considering the remaining torsion branes above, the $(3;1,0,0)$ and
$(3;2,0,0)$ branes are related by decay, and the $(3;2,2,1)$ and
$(3;2,2,2)$ branes are also related by decay. Since the $(3;2,0,0)$,
$(3;0,2,0)$, $(3;0,0,2)$, and $(3;2,2,2)$ branes are not related by
decay, then these branes form an independent basis of K-theory
torsion charge, and match with the probe brane results in eqns.
(\ref{eqn:kusual}) - (\ref{eqn:kusual4}).

\subsection{Torsion branes in the tensor-multiplet model}
\label{subsec:tortens}

In this subsection we continue our analysis to the torsion branes in
the tensor-multiplet model. For the tensor-multiplet model the
choice of discrete torsion is
\begin{equation}
\kappa=-1\,,\qquad \kappa_{\Omega R}=1\,,\qquad \kappa_{\Omega R
g_i}=-1\,,
\end{equation}
and equation~(\ref{eqn:tachbegone2}) together with the
conditions~(\ref{eqn:DPBZorient}) imply that tachyon-free torsion
D-branes of the type given in equation~(\ref{eqn:threetors}) need to
have $(r;s_1,s_2,s_3)$ of the form
\begin{equation}
(-1;2,2,2)\,,\qquad (0;2,2,2)\,,\qquad (0;0,0,0)\,,\qquad
(1;0,0,0)\,,\qquad (2;0,0,0)\,.
\end{equation}
Tachyon-free torsion D-branes of the type given in
equation~(\ref{eqn:onetors}), with couplings to the NSNST$g_1$
sector, need to have $(r;s_1,s_2,s_3)$ of the form
\begin{eqnarray}
&&(-1;2,2,2)\,,\quad (-1;1,2,2)\,,\quad (0;2,2,2)\,,\qquad
(0;1,2,2)\,,\qquad
(0;0,0,0)\,,\nonumber \\
&&(0;1,0,0)\,,\qquad (1;0,0,0)\,,\qquad (1;1,0,0)\,,\qquad
(2;0,0,0)\,,\qquad (2;1,0,0)\,.
\end{eqnarray}
Finally, tachyon-free torsion charged D-branes of the form given in
equation~(\ref{eqn:notors}) need to have $(r;s_1,s_2,s_3)$ of the
form
\begin{eqnarray}
&&(-1;2,2,2)\,,\quad (-1;1,2,2)\,,\quad (-1;1,1,2)\,,\quad
(-1;1,1,1)\,,\qquad (0;2,2,2)\,,\qquad \nonumber \\
&&(0;1,2,2)\,,\qquad (0;1,1,2)\,,\qquad (0;1,1,1)\,,\qquad
(0;0,1,1)\,,\qquad (0;0,0,1)\,,\nonumber \\
&&(0;0,0,0)\,,\qquad (1;0,0,0)\,,\qquad (1;1,0,0)\,,\qquad
(1;1,1,0)\,,\qquad (1;1,1,1)\,,\nonumber \\ &&(2;0,0,0)\,,\qquad
(2;1,0,0)\,,\qquad (2;1,1,0)\,.
\end{eqnarray}
and all permutations of the $s_i$.

It is easy to see that the $(0;0,0,0)$-brane and the
$(1;1,0,0)$-brane of the type~(\ref{eqn:onetors}) are $g_3$ images
of the ${\bf Z_2}\oplus{\bf Z_2}$ $(2,4)$-branes found
in~\cite{StQu}\footnote{The $(2,4)$-brane was not explicitly
mentioned as an inconsistent brane in \cite{StQu}. See Section
\ref{sec:GPBLcomp} for a brief review of the 6D tensor multiplet,
including the equations needed to calculate the spectrum of stable
D-branes.}. These branes were shown not to be
consistent~\cite{StQu}, and so we exclude them here as well. For
branes of the type (\ref{eqn:notors}), the $(-1;1,1,1)$,
$(0;1,1,2)$, and $(2;1,1,0)$ branes are also inconsistent and must
be removed. Since it is possible to form branes of the
type~(\ref{eqn:onetors}) from branes of the
type~(\ref{eqn:threetors}), and branes of the
type~(\ref{eqn:notors}) from branes of the type~(\ref{eqn:onetors}),
we need to exclude all torsion branes with the above values of
$(r;s_1,s_2,s_3)$.

We may next consider torsion branes of the type~(\ref{eqn:notors}).
It is possible to show that the following decay processes between
torsion-charged branes of this type can occur for suitable values of
radii
\begin{equation}
(0;1,1,1)\rightarrow(0;0,1,1)\rightarrow(0;0,0,1)\rightarrow(0;0,0,0)\,.
\end{equation}
as well as
\begin{equation}
(1;1,1,1)\rightarrow(1;1,1,0)\rightarrow(1;1,0,0)\,.
\end{equation}
Since the end of each of these decays is an inconsistent D-brane,
all of the D-branes in the above decays are also inconsistent.

To summarize, the tensor-multiplet model includes torsion D-branes
of the type given in equation~(\ref{eqn:threetors}) with
$(r;s_1,s_2,s_3)$ of the form
\begin{equation}
(-1;2,2,2)\,,\qquad (0;2,2,2)\,,\qquad (1;0,0,0)\,,\qquad
(2;0,0,0)\,.
\end{equation}
The tensor model has torsion D-branes of the type given in
equation~(\ref{eqn:onetors}), with couplings to the NSNST$g_1$
sector, with $(r;s_1,s_2,s_3)$ of the form
\begin{eqnarray}
&&(-1;2,2,2)\,,\qquad (-1;1,2,2)\,,\qquad (0;2,2,2)\,,\qquad
(0;1,2,2)\,,\\ \nonumber && (1;0,0,0)\,,\qquad (2;0,0,0)\,,\qquad
(2;1,0,0)\,.
\end{eqnarray}
Similar results hold for branes of the type given in
equation~(\ref{eqn:onetors}) with couplings to the other NSNST$g_i$
sectors. Finally, the tensor model has torsion D-branes of the form
given in equation~(\ref{eqn:notors}) for $(r;s_1,s_2,s_3)$ of the
form
\begin{eqnarray}
&&(-1;2,2,2)\,,\quad (-1;1,2,2)\,,\quad (-1;1,1,2)\,,\quad
(0;2,2,2)\,,\\ \nonumber && (0;1,2,2)\,,\qquad (1;0,0,0)\,,\quad
(2;0,0,0)\,,\qquad (2;1,0,0)\,.
\end{eqnarray}
and all permutations of the $s_i$.

\subsection{A comment on other choices of discrete torsion}
\label{subsec:distortens}

Naively, there are many torsion charged branes that are a solution
to eqn (\ref{eqn:tachbegone2}). Though in our analysis for the
hyper-multiplet model we have excluded some of the branes in Tables
\ref{tab:g1torstable} and \ref{tab:numbertorstable} due to their
relation to inconsistent branes, this \emph{does not} mean that for
other choices of discrete torsion that the discrete K-theory charges
found using the probe brane approach in equations (\ref{eqn:ktors})
- (\ref{eqn:ktors4}) are also excluded if branes that could carry
those charges are excluded. These discrete K-theory charges can be
carried by BPS branes even when non-BPS branes that also carry the
same charges do not exist. The probe brane approach in the
hyper-multiplet uses a configuration of BPS $D3$ and $D7$ branes to
detect these discrete charges, and the same charges can also be
carried by non-BPS torsion charged $D9$ and $D5$ branes
respectively. This method is independent of whether the non-BPS
torsion charged $D9$ and $D5$ branes exist. In addition, the
discussion above is based upon changing the compactification radii
of the torsion branes along certain directions so that they can
decay. By doing this we are moving the model away from its BPS
configuration, and thus must treat this analysis with caution.

There are interesting features for the torsion brane spectrum for
other choices of discrete torsion. Some of the other 14 cases
contain branes that do not fill the non-compact directions (i.e.
$r\neq 3$), and can possibly carry torsion charge not carried by the
$r=3$ branes. In addition, these models can contain $r=3$ branes
with oblique flux on their worldvolume, and might be useful for
future model building. To determine the torsion brane spectrum, a
case by case analysis must be done to determine whether the branes
are stable, orientifold and orbifold invariant, and consistent. A
full study of the remaining cases would be useful both to string
phenomenology and to the study of twisted K-theory.

\section{Some Known Examples}
\label{sec:GPBLcomp} \setcounter{equation}{0}

Here, we will examine a previously worked example of a $\bf{T}^4 /
\bf{Z}_2$ and a T-dual version of the $\bf{T}^6 / \bf{Z}_2 \times
\bf{Z}_2$ considered in this paper. The former corresponds to the
hyper-multiplet model of a $\bf{T}^4 / \bf{Z}_2$ orientifold
\cite{Pradisi:1988xd, GP} analyzed in \cite{StQu} where the
underlying K-theory group structure was also discussed. The model
contains 1 O5 and 1 O9 plane whose charge is cancelled by
introducing D5 and D9 branes. The RR tadpole conditions are
\begin{eqnarray} \frac{v_6 v_4}{16}\{32^2 - 64
{\rm{Tr}}( \gamma^{-1}_{\Omega,9}\gamma^{T}_{\Omega,9}) +
({\rm{Tr}}(\gamma_{1,9}))^2\}
\\ \nonumber
+ \frac{v_6}{16v_4} \{32^2 - 64 {\rm{Tr}}( \gamma^{-1}_{\Omega
R,5}\gamma^{T}_{\Omega R,5}) + ({\rm{Tr}}(\gamma_{1,5}))^2\} \\
\nonumber + \frac{v_6}{64} \sum_{I=1}^{16}
\{{\rm{Tr}}(\gamma_{R,9})-4 {\rm{Tr}}(\gamma_{R,I})\}^2=0
\end{eqnarray}
The open string projection operator is
\begin{eqnarray} \nonumber
(\frac{1+ \Omega }{2})(\frac{1+ I_4}{2})(\frac{1+ (-1)^F}{2})
\end{eqnarray}
Adapting the results from \cite{StQu}, the tachyon cancellation
condition\footnote{In the $\bf{T}^4$, r $\leq$ 5 and s $\leq$
4.}${}^{,}$\footnote{See Table I in \cite{StQu} for more details on
the value of $n$ for different values of $r$ and $s$.} is
\begin{eqnarray}
2^4 n^2 (1 + \epsilon)
- 2 \sqrt{2} ~n ~\sin (\frac{\pi}{4} (r+s-5))  - 2
\sqrt{2} ~ n ~\sin (\frac{\pi}{4} (r-s-1)) = 0
\end{eqnarray}
For $\epsilon=-1$ this corresponds to integrally charged non-BPS
D-branes that couple to the twisted \RR {} sector, and reproduces
the result $s$=0,4 for all $r$ and $r=-1,3$ for all $s$. The other
option of $\epsilon=1$ gives us torsion branes that couple to the
twisted NSNS sector, and are stable for $r$=5, $s$=2.

In \cite{StQu} the tensor-multiplet model was also investigated. As
mentioned previously, the tensor-multiplet differs from the
hyper-multiplet by a choice of discrete torsion. Changing the
discrete torsion between the orientifold and $\cI_4$, is equivalent
to changing the sign in the M\"{o}bius strip diagram in the $Dp$ -
$O5$ amplitude. The new tachyon cancellation condition for torsion
branes in the tensor-multiplet is
\begin{eqnarray}
\label{eqn:DPBZtors} 2^4 n^2 ( 1 + \epsilon)
-  2 \sqrt{2} ~ n ~\sin (\frac{\pi}{4} (r+s-5))  + 2
\sqrt{2} ~ n ~\sin (\frac{\pi}{4} (r-s-1)) = 0
\end{eqnarray}
The BPS fractional branes have $r,s$ = (1,0), (1,4), (5,0), and
(5,4). The new non-BPS branes from eqn. (\ref{eqn:DPBZtors}) are
$r$=1,5, $s$=1,2,3 for the integrally charged branes, and for the
torsion branes $r$=-1,0 and $s$=0 or $r$=3,4 and $s$=4.

Let us now consider the $\bf{T}^6 / \bf{Z}_2 \times \bf{Z}_2$
orientifold in the T-dual frame where there are O9-planes and 3
types of O5 planes \cite{BL}. The RR tadpole conditions are
\begin{eqnarray}
\{v_1 v_2 v_3 (32^2 - 64 {\rm{Tr}}(
\gamma^{-1}_{\Omega,9}\gamma^{T}_{\Omega,9}) +
({\rm{Tr}}(\gamma_{1,9}))^2)
\\ \nonumber
\sum_i v_i \prod_{j \neq i} \frac{1}{v_j} (32^2 - 64 {\rm{Tr}}(
\gamma^{-1}_{\Omega R_i,5_i}\gamma^{T}_{\Omega R_i,5_i}) +
({\rm{Tr}}(\gamma_{1,5_i}))^2 \}
\end{eqnarray}
The open string projection operator is
\begin{eqnarray}
(\frac{1+ \Omega}{2}) (\frac{1+ g_1}{2})(\frac{1+ g_2}{2})
\end{eqnarray}
which lead to the following tachyon cancellation condition
\begin{eqnarray}
2^4~n^2 \times (1 + \epsilon_{T_1} + \epsilon_{T_2} +
\epsilon_{T_3})
\\ \nonumber - 2n~
\sin(\frac{\pi}{4}(r+s-5)) \\ \nonumber - 2n~\sin(\frac{\pi}{4}( r-s+2s_3-1 ))\\
\nonumber - 2n~\sin(\frac{\pi}{4}(r-s+2s_1-1 ))\\
\nonumber - 2n~ \sin(\frac{\pi}{4}(r-s+2s_2-1 )) = 0
\end{eqnarray}
All of the branes we find are T-dual to the branes in Section
\ref{sec:torb}, and are related under the transformation $s_i
\longrightarrow 2 - s_i$.

\section{Other Choices of Discrete Torsion}
\label{sec:clearing}

Table \ref{tab:gaugegroup} contains the gauge group for the open
strings in the 33 and $7_i 7_i$ sectors. Only 8 of the possible 16
choices of discrete torsion have been listed; the other 8 cases can
be obtained from a permutation of our results. In addition, we have
included tables of non-BPS integrally charged branes with $r=0$,
since these are possible D-matter candidates \cite{Shiu:2003ta} in
addition to the torsion branes. The integrally charged brane
spectrum for different choices of discrete torsion is listed in
Table \ref{tab:dmatter}. As noted in Section \ref{subsec:torbrane},
a consistent open string projection requires each of these
integrally charged branes to couple to only one twisted sector. Our
results are only stated for 8 of the possible 16 choices of discrete
torsion. The spectrum for the other cases can be obtained through a
simple permutation of the branes in the tables. Branes that are
shown to be inconsistent are marked with a dagger.

\begin{table}[ht]
\begin{center}
\begin{tabular}{|c|c|c|c|c|c|c|c|c|} \hline
$\kappa$ &$\kappa_{\Omega R}$ &$\kappa_{\Omega Rg_1}$
&$\kappa_{\Omega Rg_2}$ &$\kappa_{\Omega Rg_3}$
&$D3$ &$D7_1$ &$D7_2$ &$D7_3$ \\
\hline

+ &+ &+ &+ &+ &\UUSp &\UUSp &\UUSp &\UUSp \\ \hline

+ &+ &+ &$-$ &$-$ &\SSO &\SSO &\UUSp &\UUSp \\ \hline

+ &$-$ &+ &+ &$-$ &\UUSp &\SSO &\SSO &\UUSp \\ \hline

+ &$-$ &$-$ &$-$ &$-$ &\SSO &\SSO &\SSO &\SSO \\ \hline

$-$ &$-$ &+ &+ &+ &\USp &\U &\U &\U \\ \hline

$-$ &$-$ &+ &$-$ &$-$ &\U &\SO &\U &\U \\ \hline

$-$ &+ &+ &+ &$-$ &\U &\U &\U &\USp \\ \hline

$-$ &+ &$-$ &$-$ &$-$ &\SO &\U &\U &\U \\ \hline
\end{tabular}
\caption{The gauge group for the open strings in the 33 and $7_i
7_i$ sectors. The gauge groups for the other eight choices of
discrete torsion can be obtained by a simple permutation of the
results above.} \label{tab:gaugegroup}
\end{center}
\end{table}

\begin{table}[ht]
\begin{center}
\begin{tabular}{|c|c|c|c|c|c|} \hline
$\kappa$ &$\kappa_{\Omega R}$ &$\kappa_{\Omega Rg_1}$
&$\kappa_{\Omega Rg_2}$ &$\kappa_{\Omega Rg_3}$
&Integrally Charged Branes $(r = 0, s)$        \\
\hline + &+ &+ &+ &+ &None \\ \hline
+ &+ &+ &$-$ &$-$
  &$s_1=1, s_j=1, s_k=0$ \\
  &  &  &    &    &$s_1=s_2=s_3=1$ \\
  &  &  &    &    &$s_1=1, s_j=1, s_k=2^{\dag}$ \\
\hline
+ &$-$ &+ &+   &$-$ &$s_1=1$ \\
  &    &  &    &    &$s_2=1$ \\
  &    &  &    &    &$s_1=1, s_j=1, s_k=0$ \\
  &    &  &    &    &$s_2=1, s_j=1, s_k=0$ \\
  &    &  &    &    &$s_1=s_2=s_3=1$ \\
  &    &  &    &    &$s_1=1, s_j=0, s_k=2$ \\
  &    &  &    &    &$s_2=1, s_j=0, s_k=2$ \\
  &    &  &    &    &$s_1=1, s_j=1, s_k=2^{\dag}$ \\
  &    &  &    &    &$s_1=2, s_2=s_3=1^{\dag}$ \\
  &    &  &    &    &$s_2=s_3=2, s_1=1$ \\
  &    &  &    &    &$s_1=s_3=2, s_2=1$ \\ \hline
+ &$-$ &$-$ &$-$ &$-$ &None \\ \hline
$-$ &$-$ &+ &+ &+
    &$s_i=0, s_j=1, s_k=2$ \\ \hline
$-$ &$-$ &+ &$-$ &$-$  &$s_1=1, s_j=0, s_k=2$ \\
\hline
$-$ &+ &+ &+ &$-$ &$s_1=1$ \\
    &  &  &  &    &$s_2=1$ \\
    &  &  &  &    &$s_2=s_3=2, s_1=1$ \\
    &  &  &  &    &$s_1=s_3=2, s_2=1$ \\ \hline
$-$ &+ &$-$ &$-$ &$-$ &None \\  \hline
\end{tabular}
\caption{D-matter candidates for different choices of discrete
torsion. Branes that are shown to be inconsistent are marked with a
dagger.} \label{tab:dmatter}
\end{center}
\end{table}

\end{document}